\documentclass[%
 reprint,
superscriptaddress,
nofootinbib,
 amsmath,amssymb,
 aps,
]{revtex4-2}

\usepackage{graphicx}
\usepackage{dcolumn}
\usepackage{listings}
\usepackage{tikz}
\usetikzlibrary{arrows.meta,decorations.markings}
\usepackage{verbatim}
\usepackage{bigints}

\newcommand{\EE}{\mathbb{E}}

\newcommand{\PP}{\mathcal{P}}

\newcommand{\pr}[1]{\operatorname{Pr}\{#1\}}
\newcommand{\OO}{\mathcal{O}}
\newcommand{\ex}[1]{\langle #1 \rangle}

\usepackage{bm}

\usepackage{xcolor}

\usepackage{hyperref}

\begin{document}

\preprint{APS/123-QED}

\title{Forecasting and decisions in the birth-death-suppression Markov model for wildfires}

\author{George Hulsey}
 \email{hulsey@physics.ucsb.edu}
 \affiliation{%
Department of Physics, \\
UC Santa Barbara, Santa Barbara, CA 93106
}%

\author{David L.~Alderson}

\affiliation{
 Operations Research Department, \\
 Naval Postgraduate School, Monterey, CA 93943
}

\author{Jean Carlson}%
 \email{carlson@ucsb.edu}
 \affiliation{%
Department of Physics, \\
UC Santa Barbara, Santa Barbara, CA 93106
}%

\date{\today}

\begin{abstract}
As changing climates transform the landscape of wildfire management and suppression, agencies are faced with difficult resource allocation decisions. We analyze trade-offs in temporal resource allocation using a simple but robust Markov model of a wildfire under suppression: the birth-death-suppression process. Though the model is not spatial, its stochastic nature and rich temporal structure make it broadly applicable in describing the dynamic evolution of a fire including ignition, the effect of adverse conditions, and the effect of external suppression. With strong analytical and numerical control of the probabilities of outcomes, we construct classes of processes which analogize common wildfire suppression scenarios and determine aspects of optimal suppression allocations. We model problems which include resource management in changing conditions, the effect of resource mobilization delay, and allocation under uncertainty about future events. Our results are consistent with modern resource management and suppression practices in wildland fire.

\end{abstract}

\maketitle

\section{Introduction}
Wildfires pose an increasing risk to life, property, and infrastructure \cite{Pandey_Huidobro_Lopes_Ganteaume_Ascoli_Colaco_Xanthopoulos_Giannaros_Gazzard_Boustras_etal_2023,Burke_Driscoll_Heft-Neal_Xue_Burney_Wara_2021,Brown_Hanley_Mahesh_Reed_Strenfel_Davis_Kochanski_Clements_2023}. In the United States (US), the problem of understanding and mitigating wildfire risk falls on state and federal agencies, local governments, and firefighting personnel, representing a multiscale challenge across public and private sectors. Every level of response to a fire event involves a series of operational and tactical decisions, often made under uncertainty. Decision makers integrate field experience, resource availability, weather forecasts, protection of assets and persons, and the cost of applying suppression resources into their strategy. At a high level,  disaster response agencies are faced with difficult decisions related to the deployment, allocation, and retrieval of fire suppression resources \cite{MaryTaber_Elenz_Langowski_2013}. 

\begin{figure}[!h]
    \centering
    \includegraphics[width=\columnwidth]{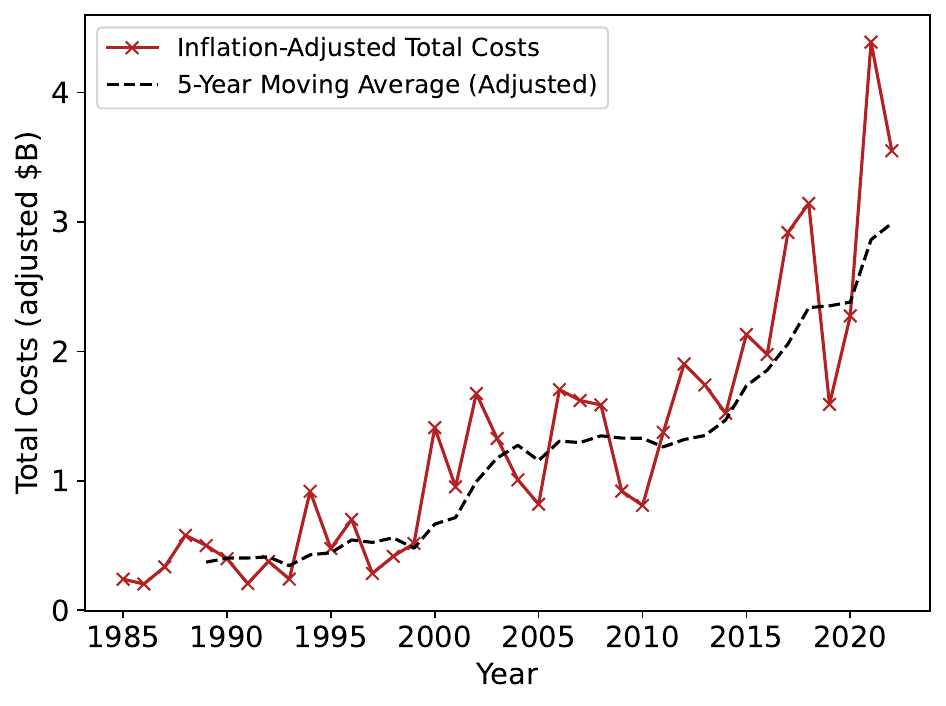}
    \caption{Gross federal expenditure (\$B), adjusted by yearly CPI, of the US Forest Service and Dept. of Interior on wildfire suppression from 1985-2023 \cite{nifc_costs}. The five-year moving average (dashed) shows growth in suppression spending which has accelerated in recent years. }
    \label{fig:cost_over_time}
\end{figure}

Changing climates across the globe have contributed to increased prevalence and severity of wildfire events over the last few decades \cite{jolly2015climate,westerling2008climate,liu2010trends}, and the anthropogenic origin of some of these effects is likely to worsen in the near future \cite{abatzoglou2016impact}. The cost of these wildfires, both in their damage and the cost of their suppression, has also increased \cite{gorte2013rising,mattioli2022estimating,bayham2022economics} as human populations grow in fire-prone areas, expanding the wildland-urban interface (WUI). Data from the National Interagency Fire Center \cite{nifc_costs} show that US federal firefighting costs, combined between the United States Forest Service (USFS) and Department of Interior (DOI), have been steadily increasing, as depicted in Fig.~\ref{fig:cost_over_time}. 

Particularly in California, shifting rain patterns have exacerbated the risk posed by autumnal offshore wind events \cite{Swain_2021}. Low fuel moisture and strong katabatic winds pose extreme fire danger late in the season \cite{Garner_Kovacik_2023,Guirguis_2023}. In the Western US, downslope winds driven by hot inland conditions---such as Santa Ana, Diablo, and others--are associated with numerous historical extreme fire events \cite{Abatzoglou_Kolden_Williams_Sadegh_Balch_Hall_2023}. These offshore flows have been analyzed as a driver and even predictor of wind-driven fires in the brushy, chaparral ecosystem characteristic of Mediterranean climates like Southern California \cite{Billmire_French_Loboda_Owen_Tyner_2014,Moritz_Moody_Krawchuk_Hughes_Hall_2010,Westerling_Cayan_Brown_Hall_Riddle_2004}. If a wind event is forecasted, management agencies must both prepare for potential ignitions as well as keep active fire events under control. Recent projections indicate that not only are fires likely to increase in severity, but the likelihood of simultaneous large fire events will also increase \cite{McGinnis_Kessenich_Mearns_Cullen_Podschwit_Bukovsky_2023}. 

Wildfire management encompasses multiple practices from ignition mitigation to active suppression \cite{prescribed_2023}. This work focuses on suppression-oriented management, where suppression resources are deployed during, after, or in anticipation of a fire event. An increasingly common practice in wildland fire management is \textit{pre-positioning}, where suppression resources are preemptively deployed in high-risk areas in order to reduce response time to a potential ignition. The National Interagency Coordination Center (NICC) provides predictive analytics, as well as resource management tools, to assist cooperation between agencies like USFS and the US DOI in their pre-positioning efforts \cite{NationalInteragencyFireCenter_2023}. Optimization-based strategies for pre-positioning assets have been directly addressed in the literature in recent years \cite{arnette2019risk,seeberger2020new,suarez2017stochastic}. 

Whereas pre-positioning is relevant before a fire event begins, the question of resource removal arises towards the end of a fire event: \emph{after adverse conditions have subsided, when is it appropriate to remove suppression resources from a fire?} The question of resource removal is crucial both for cost management and maintaining the ability to respond to other potential events. Optimizing resource removal has received significantly less attention in the disaster response literature than pre-positioning. 

The resource allocation decisions associated with wildfire suppression can be quantitatively addressed through a simple but robust stochastic model: the birth-death-suppression Markov process. The model is extremely general and describes the temporal evolution of a fire, taking a mean-field theory approach to the spatial fire dynamics. The model addresses temporal trade-offs: when to apply suppression, when to remove it, and how strategies are affected in the face of forecasted changes in conditions or uncertainty about the future. In contrast, the spatiotemporal evolution of a wildfire is a very difficult modeling problem. By considering only the temporal dimension, the analysis is designed to be robust to detailed and unpredictable small-scale spatial variations.

\section{The birth-death-suppression process} 
\label{sec:bds_general}
Introduced in \cite{petrovic}, the birth-death-suppression process is a linear Markov birth-death process that describes the stochastic evolution of an abstract population $j(t)$. In the present context, the population $j(t)$ represents the number of actively burning spatial units of fire (e.g., acres), which we refer to as \textit{firelets}. While our focus here is describing the size of a wildfire, this process, and birth-death processes more generally, can also be used to model the dynamics of epidemics or other human population-oriented phenomena \cite{kendall1949stochastic}. 

The model is temporal, with transitions (births and deaths) $j \to j \pm 1$ occurring over time.  These births and deaths represent the ignition and extinction of individual firelets. The time evolution of the model depends on three parameters: the birth rate $\beta$, characterizing the number of new ignitions per firelet per unit time, the death rate $\delta$, characterizing the number of natural extinctions per firelet per unit time, and the suppression rate $\gamma$, representing the effect of external suppression on the fire. For a population of size $j(t)$, births occur on aggregate at a rate $\lambda_j = \beta \cdot j(t)$ and deaths occur on aggregate at a rate $\mu_j = \delta  \cdot j(t) + \gamma$, per unit time. In what follows, we explicitly set the death rate $\delta = 1$. This corresponds to choosing units of time such that the interval $\Delta t = 1$ is the average time until a single firelet extinguishes due to fuel exhaustion. 

In previous work \cite{bds_wildfire}, the theory of the model is studied in detail; explicit analytical expressions for the transition probabilities and asymptotic outcomes of the process are developed. The birth-death-supppression process and related Markov processes have been studied by numerous authors in the mathematical literature \cite{karlinear,kendallGeneral,karlinDiff,karlinClass,randomwalksKarlin}; for more, see \cite{bds_wildfire} and the references therein. 

The Markov process is absorbing at zero, in the sense that processes terminate once the population reaches the state $j(t) = 0$; this represents the complete extinguishing of a fire. The process is transient: populations either eventually absorb or diverge in size. However, considering only the actively burning size of the fire neglects the cumulative nature of the burned area of the fire. To track the total burned area (the cumulative number of ignitions), one must consider the \textit{footprint} $F(t)$, defined as the variable which shares all births of the population: transitions occur jointly as $(j \to j+1,\ F\to F+1)$ and $(j \to j -1,\ F \to F)$. 

The footprint represents the cumulative spatial extent of the fire. A principal outcome of interest is \textit{escape}, defined here as the footprint exceeding some threshold $F(t) \geq J$. Escape represents the fire growing above a reasonable containment size or growing large enough to impact built infrastructure. The escape probability $\pr{F \geq J}$ is a diagnostic of the riskiness of the process and is a focus of much of the analytics in this work. Integral expressions exist for the asymptotic distribution of footprints, and hence the escape probability, for a process with some initial size and arbitrary parameters $\beta, \gamma \geq 0$. The mathematical details of these formulae are reviewed in Appendix \ref{sec:app_glossary}, and a more detailed exposition of the birth-death-suppression process is available in \cite{bds_wildfire}. 

The goal of this paper is to quantitatively address common scenarios in wildfire resource management by modeling them with the birth-death-suppression process, over which we have strong analytical and numerical control. The focus is on problems of resource management which arise with variable conditions, where the birth rate $\beta$ of the process changes in time, modeling the drastic changes in fire conditions that can be brought on by high wind events. The benefits of resource pre-positioning are considered by studying the effect of suppression delay on outcomes of the process, as are related questions of suppression allocation under uncertainty about future fire events. The results are consistent with the conventional and practical wisdom of fire suppression, specifically, the importance of `initial attack' and the concentration of suppression resources. 

\textbf{The dynamics of the process.} 
The birth-death-suppression process fundamentally describes the time evolution of the population $j(t)$ from some initial size $j(0) \equiv N$. As the population grows, births occur at a rate proportional to its size, and deaths occur by two mechanisms: natural extinction, proportional to $j(t)$, and external suppression, proportional to the suppression rate $\gamma$. In the absence of suppression ($\gamma = 0$), on average, the population experiences either exponential growth or decay:
\begin{equation}
    \ex{j(t)} = Ne^{(\beta-1)t}.
\end{equation}
The footprint, which counts all births of the population, either grows exponentially with the population (in the case $\beta > 1$) or saturates at an asymptotic value (in the case $\beta < 1$). It is generally true, for any $\gamma$, that the average footprint $\ex{F(t)}$ solves the differential equation
\begin{equation}
    \label{eq:change_in_footprint}
    \frac{d\ex{F(t)}}{dt} = \beta\ex{j(t)}, 
\end{equation}
which reflects the fact that the average footprint grows exactly as the average aggregate birth rate of the process $\ex{\lambda_j} = \beta \ex{j(t)}$. Simple integration gives the average footprint with zero suppression as
\begin{equation}
\langle F(t)\rangle=\frac{N}{\beta-1}\left(\beta e^{(\beta-1) t}-1\right).
\end{equation}
The critical point of the process occurs when births and deaths happen with equal probability $(\beta = 1)$. In \cite{bds_wildfire}, an explicit expression for the asymptotic escape probability $\pr{F_\infty \geq J}$ in the case $\beta = 1,\ N = 1$ is found: 
\begin{equation}
\pr{F_\infty \geq J} = \frac{\Gamma(J-1 / 2) \Gamma(1+\gamma / 2)}{\sqrt{\pi} \Gamma(J+\gamma / 2)}.
\end{equation}
Asymptotically in the escape threshold $J$, this probability is a power law:
\begin{equation}
    \label{eq:asym_escape_exact}
   \pr{F_\infty \geq J}  \sim \frac{\Gamma\left(1+\frac{\gamma}{2}\right)}{\sqrt{\pi J^{1+\gamma}}}+\mathcal{O}(J^{-3 / 2}).
\end{equation}
One motivation for the use of this model to describe wildfire is the empirical distribution of wildfire footprints (burned areas), which is known to be approximately power-law distributed $P(F\geq J) \sim J^{-\alpha}$ with exponent $\alpha \approx 1/2$ \cite{Moritz_Morais_Summerell_Carlson_Doyle_2005}. The same distribution is found in the birth-death-suppression model near the critical point, as in Eq.~\eqref{eq:asym_escape_exact} where $\alpha = 1/2 + \gamma/2$. 

Analytical formulae for asymptotic probabilities like Eq.~\eqref{eq:asym_escape_exact} when $\beta \neq 1$ can be found by the method of orthogonal polynomials \cite{karlinear}, but in general, they are expressed as highly complex integrals which must be evaluated numerically. In order to study the dynamics of the process one must therefore fix numerical values of the parameters. 

\textbf{Choice of numerical parameter scales.} The most general birth-death-suppression process includes a death rate $\delta$ such that $\lambda_j = \beta j,\ \mu_j = \delta j + \gamma$. We set $\delta = 1$, or correspondingly choose units of time $t$ such that the interval $\Delta t = 1$ is the average time-to-death of a single unit of the population\footnote{To restore the death rate $\delta$ to any formulae, one simply makes the replacements $t\to \delta t,\ \beta \to \beta/\delta,\ \gamma \to \gamma/\delta$. Any product $\beta t,\ \gamma t$ is therefore unaffected by setting $\delta = 1$.}. In the fire interpretation, this means one unit of time is the average time that, for example, an acre of land burns before naturally extinguishing due to fuel exhaustion. Converting the duration $\Delta t = 1$ to physical time units, i.e. minutes, is roughly equivalent to specifying the spatial extent of one ``unit" of the population: an acre, a square kilometer, etc. Timescales are fixed to values $T \lesssim 10$ for numerical simplicity. Thus, if one takes a firelet to be a single acre, then here we consider timescales equivalent to the average time it takes 10 acres to naturally extinguish. 

Numerical values of the birth rate are chosen which are close to $\beta = 1 ( = \delta)$, where the statistics of the footprint distribution match the empirically observed statistics of wildfire sizes. Numerical values of the suppression rate $\gamma$ implicitly represent multiples of the death rate: $\gamma = 2$ means that suppression of the fire results in the entire population being reduced by 2 firelets in the time it takes one firelet to naturally extinguish. 

Finally, we often use an initial size of $j(0) = 10$ for the population. This choice ensures that the median process evolves for a time consistent with the specified timescales $T \lesssim 10$. When $N,\gamma \gg 1$, the median lifetime of a process is approximately
\begin{equation}
    T_m \approx \frac{1}{1-\beta} \log \left[1+\frac{N}{\gamma}(1-\beta)\right],
\end{equation}
which in the critical limit $\beta \to 1$ reduces to $T_m \approx N/\gamma$. Therefore, with the numerical timescales, an initial size $N = 10$ and values of the suppression rate $\gamma \lesssim 10$ are all numerically consistent and result in dynamics which are generally bounded but still include a variety of behaviors and outcomes. 

\section{A high wind scenario}
\label{sec:multi_stage}
This paper focuses on the structure of time-dependent trade-offs in suppression allocation for a birth-death-suppression process. These trade-offs are studied through a multi-stage scenario with changing birth rates over time and a discrete set of decision epochs where a given suppression rate $\gamma$ can be applied to control the process. 

\begin{figure}
    \centering
    \includegraphics[width = \columnwidth]{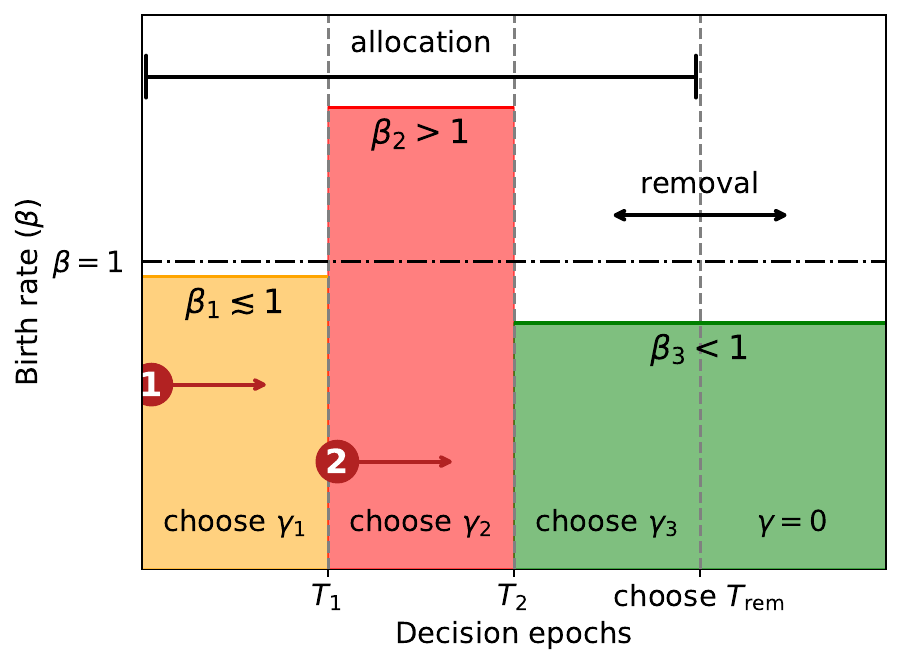}
    \caption{A schematic illustration of the high wind scenario. The scenario begins with moderate conditions which become dangerous at time $T_1$. Later, at time $T_2$, the conditions relax.  We consider two ignition cases. Case 1 starts with an ignition in advance of a forecasted high-wind event, and Case 2 starts with an ignition during the high-wind event.  A strategy consists of a suppression allocation $(\gamma_1,\gamma_2,\gamma_3)$ for each epoch in addition to a time $T_{rem}$ after which no more suppression is applied. }
    \label{fig:high_wind_schematic}
\end{figure}

With problems of resource allocation that arise in wildfire management in mind, a `high wind scenario' birth-death-suppression process is constructed.  In this setup, a process begins at $T = 0$ with a moderate, but not supercritical\footnote{For any $\gamma$, we refer to the phase $\beta > 1$ as \textit{supercritical}, the $\beta = 1$ point as \textit{critical}, and the $\beta < 1$ phase \textit{subcritical}. Only in the supercritical phase is asymptotic absorption at the state $j = 0$ not a certainty.}, birth rate ($\beta \lesssim 1$). After some finite time $T_1$, a forecast calls for the worsening of conditions, which is interpreted as the onset of a high wind event. For some time, the conditions are dangerous, and the birth rate is increased ($\beta > 1$) so that the process is in a supercritical regime, representing the effect of high winds on the spread of the fire. After another finite interval, at time $T_2$, the conditions relax, and the birth rate lowers, moving the process back into a subcritical regime ($\beta < 1$). A schematic of this multi-stage event is shown in Fig.~\ref{fig:high_wind_schematic}. The aim of this scenario is to model a temporally localized wind event. As a particular example, katabatic wind events vary in their duration but can be temporally localized, as is the case with `sundowner' winds observed on the California coast \cite{Ryan_1991}.

We are interested in understanding optimal strategies and trade-offs for effective management of this scenario, that is, prevention of undesirable outcomes such as the fire reaching some specified total size. Here, a strategy is defined by a choice of suppression rate $(\gamma_1,\gamma_2,\gamma_3)$ for each interval shown in Fig.~\ref{fig:high_wind_schematic} in addition to a choice of removal time $T_{rem}>T_2$ after which no suppression is applied $(\gamma = 0)$. By zero suppression, we do not literally mean a complete absence of fire suppression resources. Instead $\gamma = 0$ is a proxy for a large attenuation in suppression effort. Since the final phase of the process is subcritical, the process is guaranteed to end in a finite time regardless of the suppression applied. 

The scenario is meant to highlight multiple tensions in the choices of how to allocate suppression to a dynamically evolving process. Broadly, with a nonzero cost for suppression resources, one always has a trade-off between the cost of suppression and the cost of undesirable outcomes. However, due to the differing conditions across the event, one also must consider when to send a given amount of resources, especially if only a finite amount of suppression is available. 

Because the parameters $\gamma_i$ are suppression \textit{rates}, the total amount of suppression resources used over a time interval $\Delta T$ is proportional to $\gamma \Delta T$. Therefore, another tension arises between fast and aggressive strategies ($\gamma \gg 1,\ T \ll 1$) or slow and sustained strategies ($\gamma \ll 1,\ T \gg 1$), each of which may use the same gross amount of resources. 

Finally, near the end of the process, the parameter $T_{rem}$ represents when suppression resources may be removed. This is an increasingly relevant question in wildland fire management. Failing to remove suppression at an appropriate time consumes resources that could be better spent, with higher risk reduction, on other existing or potential events. Despite this, even fires which on average are shrinking in active size can be dangerous if they are in close proximity to infrastructure and persons. 

\subsection{Simulating the process}
To better understand the behavior of the multi-stage process we fix some of the parameters and observe simulated outcomes. There are three decision epochs each with an associated birth rate $\beta_i$ and duration $\Delta T_i$, and in each epoch one may choose a suppression rate $\gamma_i$.

The process begins at $T=0$ with initial size $j(0) = F(0) = 10$ and with birth rates constant over specified time intervals according to Table \ref{tab:param_choices}. 
\begin{table}[!h]
    \centering
    \begin{tabular}{c||c|c|c}
       Stage  &  1 (before) & 2 (wind event) & 3 (after) \\
       \hline
        Birth rate $\beta_i$ & 0.95 & 1.5 & 0.9\\
        Duration $\Delta T_i$ & $3 = T_1$ & $3 = T_2 -T_1$ & $T_{rem} - T_2$
    \end{tabular}
    \caption{Parameter choices used for simulation in this section. Here we use $j(0) = F(0) = 10$.}
    \label{tab:param_choices}
\end{table}

In the following, an ensemble of 5000 multi-stage processes is simulated, each described by the schematic of Figure~\ref{fig:high_wind_schematic} and with parameters given by Table \ref{tab:param_choices}. To characterize the evolution of outcomes, four quantities of interest are recorded: the absorption probability $p_A(t)$, the escape probability $E_J(t) = \pr{F(t) \geq J}$ (shown here with $J = 50$), and the average population and footprint $\langle j(t)\rangle,\ \langle F(t)\rangle$. 
For the latter two quantities, we also record the interquartile range of their empirical distributions and show these as intervals about the mean. 

\begin{figure}[!h]
    \centering
    \includegraphics[width = \columnwidth]{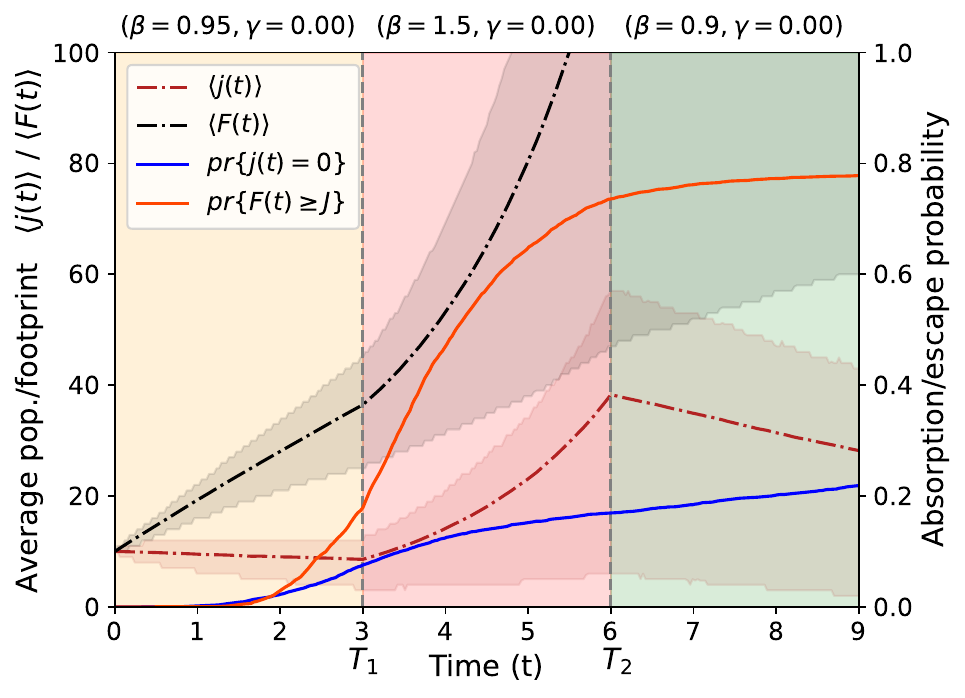}
    \caption{Zero Suppression.  Outcomes for an ensemble of 5000 simulated processes of the high wind scenario with zero applied suppression. The average population $\ex{j(t)}$ and footprint $\ex{F(t)}$ are shown on the left axis, and the probability of absorption $\pr{j(t) = 0}$ and escape $\pr{F(t) \geq J}$ are shown on the right axis. Each decision epoch is shaded corresponding to the schematic of Fig.~\ref{fig:high_wind_schematic}.
    The probability of absorption $\pr{j(t) = 0}$ (solid blue) remains low while the probability of escape $\pr{F(t) \geq J}$  (solid red) climbs quickly. The effect of the high birth rate (during $T_1 \leq t \leq T_2$) on the average population and footprint $\ex{j(t)},\ex{F(t)}$ is apparent.}
    \label{fig:zero_g_simulated}
\end{figure}

Figure~\ref{fig:zero_g_simulated} shows a simulated process with zero suppression. The dynamics introduced in Section \ref{sec:bds_general} are clearly visible. The average population (red, dash-dot, with quartiles shaded) exhibits exponential growth or decay dependent on the birth rate $\beta$ in each stage. Similarly, the onset of of the wind event in the second stage, $T_1 \leq t \leq T_2$, causes the average footprint (black, dash-dot, with quartiles shaded) to explode in size. This is consistent with the behavior of modestly sized fires after the onset of high wind conditions. The escape probability $\pr{F(t) \geq J}$ (orange-red, solid) is always increasing and saturates by the final stage, where the footprint of a majority of processes exceeds the threshold $J = 50$. The probability of absorption (blue, solid) remains small throughout, reflecting that in the absence of suppression, the wind event has made it quite unlikely that any given process will terminate on its own before the end of the timeframe pictured. 

As an initial investigation, it is instructive to compare the outcomes of the unsuppressed process to simulations in which different suppression strategies are applied. 
Two such strategies are considered: `constant' and `preemptive.' Here, we use preemptive to refer to suppression in the first epoch (yellow in Fig.~\ref{fig:zero_g_simulated}), before the onset of the high wind (red in Fig.~\ref{fig:zero_g_simulated}). 
To meaningfully compare the outcomes of differing strategies one should fix some maximal amount of suppression resources $\sum_i \gamma_i\Delta T_i$. Assuming $T_{rem} = 9$, we set $\sum_i \gamma_i\Delta T_i = 15$. 

\begin{figure}[!h]
    \centering
\includegraphics[width = \columnwidth]{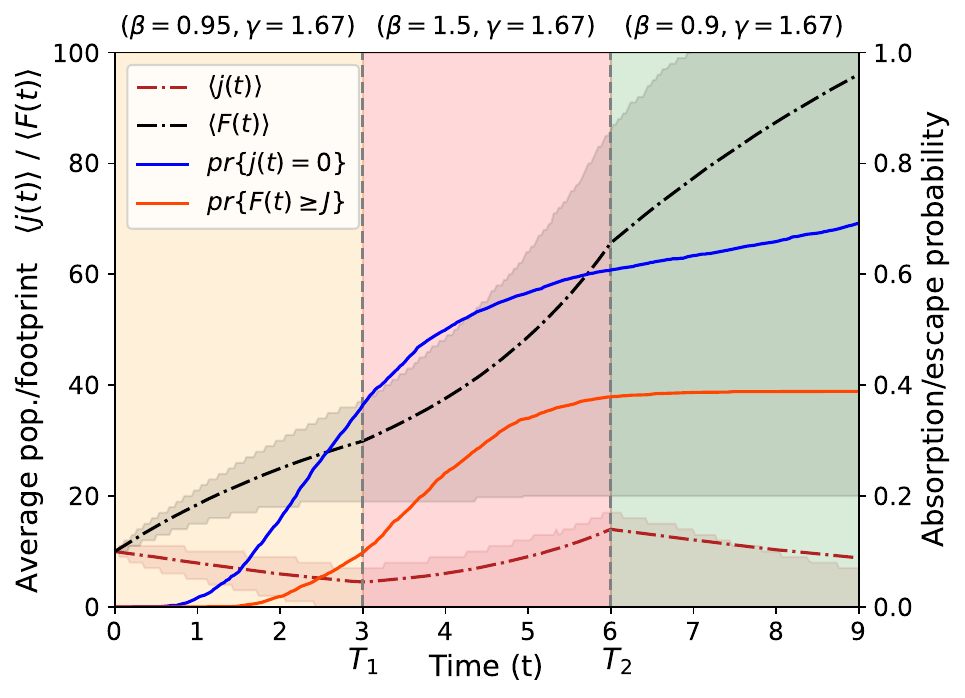}
    \caption{Constant Suppression. Outcomes for a simulated ensemble with a constant suppression schedule applied. The probability of absorption $\pr{j(t) = 0}$ (solid blue, shown on right axis) is now larger than the probability of escape $\pr{F(t)\geq J}$. The growth of the average footprint $\ex{F(t)}$ (dashed, black) is closer to linear, in stark contrast to the unsuppressed results.}
    \label{fig:consistent_simul}
\end{figure}

First, define the constant strategy as $(\gamma_1,\gamma_2,\gamma_3) = (5/3,5/3,5/3)$: a constant total amount of suppression $\gamma_i\Delta T_i$ is applied in each stage, $i=1,2,3$. Simulated outcomes using this suppression strategy are shown in Fig.~\ref{fig:consistent_simul}. 
With steady suppression applied, the growth of the footprint is severely blunted and the probability of escape $F\geq 50$ drops to just under 40\%, having been almost 80\% at the end of the un-suppressed process. 
\begin{figure}[!h]
    \centering
\includegraphics[width = \columnwidth]{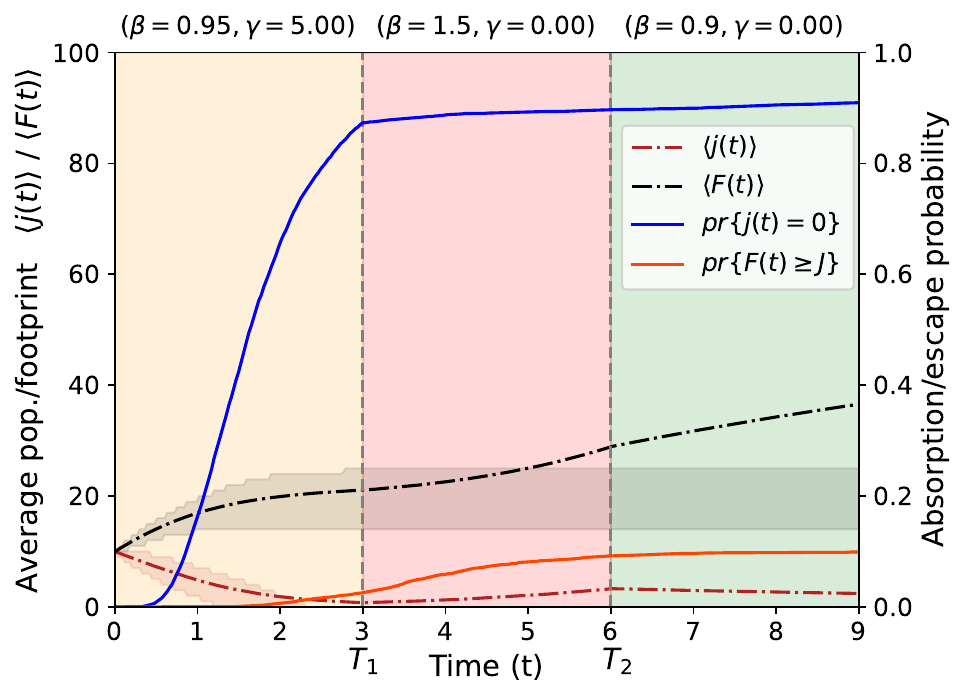}
    \caption{Preemptive Suppression. Outcomes for a simulated ensemble with a preemptive suppression schedule applied. While using the same gross amount of suppression, the higher suppression rate before the process enters dangerous conditions during times $T_1 \leq t \leq T_2$ is extremely effective at reducing the probability of escape $pr{F(t) \geq J}$ (solid red) and keeping the size of the process $\ex{F(t)}$ (dashed, black) extremely small in aggregate.}
    \label{fig:preemptive_simul}
\end{figure}

Clearly, the presence of suppression has drastically altered the outcomes. But how does this constant suppression strategy compare to a preemptive strategy, at least with the parameter choices made here? In particular, we consider allocating all available suppression to the first stage of the process, defining the preemptive strategy by $(\gamma_1,\gamma_2,\gamma_3) = (5,0,0)$ so that the total suppression resource usage is identical in both strategies. Simulated results of this strategy are shown in Fig.~\ref{fig:preemptive_simul}.

The preemptive strategy is clearly more effective in the specified parameter regime. By using more resources early, the effect of the worsening conditions in the middle epoch has been blunted. However, preemptive attack is not always feasible or guaranteed to be optimal. There may be mobilization constraints or a lack of knowledge about future conditions. If the wind event is not expected in advance or its occurrence is deemed unlikely, there may less reason for preemptive suppression. The problem of allocation under uncertainty about future events is addressed in Sec. \ref{sec:pure_allocation}.  

In both Figs. \ref{fig:consistent_simul}, \ref{fig:preemptive_simul}, the interquartile ranges do not necessarily include the average population or average footprint curves, and there is a natural reason for this. The footprint, counting the births of the process, is a sum of many Poisson processes\footnote{In a given state, births occur as a Poisson process, in the sense that the waiting time until a transition is exponentially distributed.} with differing scales, and is approximately gamma-distributed at any given time. Its support at large values/slow asymptotic decay gives the distribution of footprints excess kurtosis. Heavy-tailed statistics are one reason these models are of interest for wildfire risk, where a minority of large events may be associated with the majority of costs. A similar effect is observed for the population distribution. While many processes absorb at $j = 0$, those that do not absorb tend to grow increasingly large at a rate increasing with their size. Therefore, while the support of the population distribution is mostly concentrated near zero, the presence of large outliers at high values brings the mean population/footprint value up, even above, the interquartile range. 

\subsection{Forecasting outcomes in the birth-death-suppression process}
To forecast outcomes, and thereby compare the effectiveness of different suppression strategies, we use a composite numerical and analytical approach. With changing conditions over time, analytically forecasting the joint evolution of the population $j(t)$ and footprint $F(t)$ is  beyond the reach of closed-form solution. While ensemble simulation of the process is one possibility for numerically estimating the joint distribution, it is an inefficient and noisy way to compute probabilities. Instead, a better method is a discretization of the first-order differential equation obeyed by the joint probability matrix $\PP_{j,F}(t)$, defined as
\begin{equation}
    \PP_{j,F}(t) = \pr{j(t) = j,\ F(t) = F}.
\end{equation}
For aggregate birth, death rates $\lambda_j,\mu_j$, possibly with time dependence, one has the differential-difference equation
\begin{multline}
    \label{eq:jointProb_ode}
    \frac{d}{dt}\mathcal{P}_{j, F}(t)=\lambda_{j-1} \mathcal{P}_{j-1, F-1}(t)+\mu_{j+1} \mathcal{P}_{j+1, F}(t)\\-\left(\lambda_j+\mu_j\right) \mathcal{P}_{j, F}(t).
\end{multline}
In the present model, the aggregate birth and death rates are $\lambda_j = \beta(t)j(t)$ and $\mu_j = j(t) + \gamma(t)$ along with the boundary conditions $\lambda_{-1} = \mu_0 = 0$. The characteristic timescale of transitions is $\sim 1/(\lambda_j + \mu_j)$. For $\OO(1)$ parameters $\beta, \gamma$, transitions tend to occur on timescales of order $\sim 1/j(t)$. A high-fidelity discretization timestep should be dominated by this scale:  $\Delta t \ll 1/j(t)$. If we are considering processes with $j \sim \OO(10)$, then choosing $\Delta t \lesssim 10^{-3}$ will be sufficiently high-resolution to be numerically stable and faithfully represent the evolution of the process defined by Eq.~\eqref{eq:jointProb_ode}. 

With a discretization timestep $\Delta t$, one truncates the state space to $j \leq j_{max},\ F \leq F_{max}$, and the differential-difference equation becomes a finite-dimensional array update in the natural way: 
\begin{multline}
    \label{eq:jointProbdiffdiff}
    \mathcal{P}_{j, F}(t+\Delta t)=\left(1-\Delta t(\lambda_j+\mu_j)\right) \mathcal{P}_{j, F}(t)\\+(\lambda_{j-1}\Delta t) \mathcal{P}_{j-1, F-1}(t)+(\mu_{j+1} \Delta t )\mathcal{P}_{j+1, F}(t).
\end{multline}
Here, the changing conditions of the multistage process are captured in the time-dependence of $\beta(t),\gamma(t)$. This discretized array update is implemented numerically to forecast finite-time outcomes. Once computed, the matrix $\PP(t)$ can be marginalized to obtain the distributions over states of $j(t),F(t)$ individually, and from there to determine any statistic of interest. The numerical work was done in compiled Python using \verb|numba| \cite{lam2015numba}, which gave huge speedups in computation over base Python loops. 

However, this discretization approach is not well-suited to forecasting the asymptotic outcomes. After the time $T_{rem}$ when suppression is removed, the process in the high wind scenario evolves freely towards eventual absorption. To compute the asymptotic probability of escape, one first uses the numerical approach to determine $\PP(T_{rem})$. Then, one makes the decomposition
\begin{multline}
    \label{eq:asym_esc_recipe}
    \pr{F_\infty  = F} = \sum_{j,F}\pr{F_\infty = F|j(0) = j,\ F(0) = F} \\\times \PP_{j,F}(T_{rem}),
\end{multline}
and computes the CDF of this distribution to find the escape probability as a function of escape threshold $J$.
Analytical formulae exist for the exact computation of the asymptotic probability in the summand for arbitrary $\beta,\ \gamma$. They are included for reference in  Appendix \ref{sec:app_glossary}, Eq.~\eqref{eq:asymp_ftpt}, along with a brief review of the mathematical objects required to construct and evaluate them. 

The forecasting strategy for analyzing outcomes in the high wind scenario is a combination of the numerical discretization of the exact joint probability matrix at finite times and the analytical expressions for the asymptotic states to forecast the final state of the process. These two approaches work efficiently in their respective domains and when combined allow reliable analysis of the evolution of the process over time.

\subsection{Decomposing the scenario}
Assuming one has suitably defined a utility function encompassing both the cost of suppression resources and the cost of undesirable outcomes, optimal strategies could be investigated by simulating many processes and finding those which maximize utility. A more refined analysis may use some kind of value or policy iteration approach in lieu of a brute-force search across strategies, and hope to converge to a utility-optimal choice. In this work we take a more systematic approach. 

To study the high wind scenario, we break it into parts, each of which represents a different aspect of the suppression allocation decision process. 

\subsubsection*{Allocation optimization}
The first part of the scenario is the onset of the high wind event, where the moderate conditions in the first stage $0 \leq t \leq T_1$ worsen when the high winds occur during the interval $T_1 \leq t \leq T_2$. An allocation $(\gamma_1,\gamma_2)$ must be made to the first two stages after an ignition occurs at time $T = 0$; this is Case 1 in Fig.~\ref{fig:high_wind_schematic}. This defines an optimal strategy subproblem we call the `allocation' problem, addressed in Section \ref{sec:pure_allocation}; we temporarily ignore the latter part of the process at times $t > T_2$. Simple cost functions are constructed and optimal strategies are analyzed which focus suppression either before or during the wind event. A related problem of suppression allocation under uncertainty about potential new fire events/ignitions is also studied.

\subsubsection*{Suppression removal}
The second problem concerns the end of the process from time $T_2$ onwards, labeled `removal' in Fig.~\ref{fig:high_wind_schematic}. Here, a strategy is a choice of suppression rate $\gamma_3$ in addition to a choice of duration, or equivalently a removal time $T_{rem}$. This subproblem is addressed in Section \ref{sec:removal}. In the removal problem, the trade-off of interest is between aggressive strategies, characterized by high suppression rates over short times, versus sustained strategies, with low suppression rates over long times. 

\subsubsection*{High wind ignitions}
Finally, we study a related problem, that of ignitions in dangerous conditions. In the multi-stage scenario of Fig.~\ref{fig:high_wind_schematic}, Case 1 represents an ignition in moderate conditions, while Case 2 represents an ignition in dangerous ($\beta >1$) conditions. Often, high wind events are responsible both for ignitions and the rapid growth or spread of the fire quickly thereafter. In Sec. \ref{sec:windignit}, a scenario is constructed where ignitions occur during dangerous conditions which subsequently relax. We address both the problem of pre-positioning, studying the effect of resource delay on outcomes, and removal of resources after the wind event has ended. 

These subproblems are approached independently; the different types of strategies required in each subproblem motivate different types of analysis. The results are combined to make statements about optimal strategies for the high wind scenario as a whole, considering for context conventional wisdom and practice in the management of disaster resources.

\section{The allocation problem}
\label{sec:pure_allocation}

The allocation problem is the choice of a suppression allocation before and during the high-wind stage. That is, one chooses suppression rates
$(\gamma_1,\gamma_2)$ for the scenario of Fig.~\ref{fig:high_wind_schematic} up to time $T_2$. Allocations of suppression resources balance the cost of the resources, their ability to be deployed, and the effect they have on a fire event. 
The risk associated with the state of the process at time $T_2$ can be quantified by the  probability of certain undesirable outcomes at time $T_2$.
The first such quantity is the probability of \textit{no absorption} by time $T_2$, which is simply
\begin{equation}
    \pr{\text{no absorption}} = 1 - p_A(T_2),
\end{equation} where $p_A(t)$ is the probability that $j(t) = 0$. The second quantity is the probability of escape by the end of the wind event: 
\begin{equation}
    \label{eq:t2escape}
    \pr{\text{escape by $T_2$}} = \pr{F(T_2) \geq J}.
\end{equation}
Both the probability of no absorption and the probability of escape are lowered by the introduction of suppression; at fixed parameters $N,\beta$ and $J \gg 1$, the asymptotic probability of escape is exponentially decreasing in the suppression rate $\gamma$. 
\subsection{Optimizing allocation}
In the following, we work with parameters as given in Table \ref{tab:param_choices} for the scenario of Fig.~\ref{fig:high_wind_schematic} up to time $T_2$ to numerically study the optimal allocations.

\begin{figure}[!h]
    \centering
    \includegraphics[width=\columnwidth]{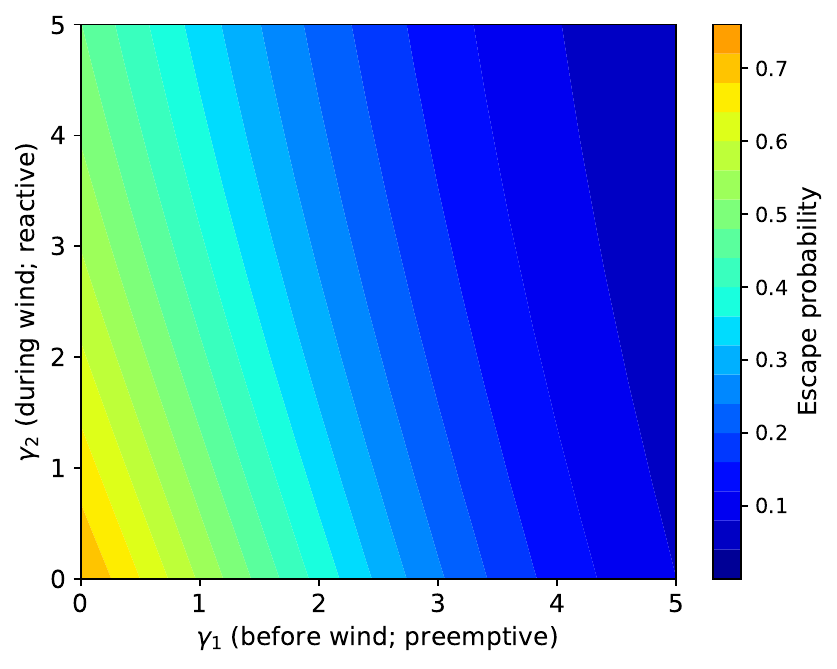}
    \caption{Effect of preemptive and reactive suppression on the escape probability. Contours of escape probability $\pr{F(T_2) \geq J}$ at time $T_2$ with a threshold $J = 50$; this represents the probability of escape by the end of the second stage in the scenario of Fig.~\ref{fig:high_wind_schematic}. In this case, preemptive suppression $(\gamma_1 >0,\ \gamma_2 = 0)$ is clearly more effective than reactive suppression $(\gamma_1 = 0,\ \gamma_2 > 0)$ for a fixed amount of resources. }
    \label{fig:alloc_esc_contour}
\end{figure}

Figure~\ref{fig:alloc_esc_contour} shows the contours of the escape probability at time $T_2$, as in Eq.~\eqref{eq:t2escape}, as a function of the preemptive suppression rate $\gamma_1$ (occurring before the wind event, $t \leq T_1$) and the reactive suppression rate $\gamma_2$ (occurring during the wind event, $T_1 \leq t \leq T_2$). 

It is clear that preemptive suppression ($\gamma_2 = 0$) has a greater effect on the probability of escape than reactive suppression ($\gamma_1 = 0$). This is not surprising: the (cumulative) footprint always increases in time, and so does the probability of escape. Waiting to apply a given amount of suppression therefore can only increase the probability of escape, regardless of the other parameters of the process. 

Some reasons that an optimal allocation would not prefer exclusively preemptive suppression are uncertainty about future conditions and/or constraints. Constraints may include limits on the ability to deploy preemptive suppression or cost constraints. Cost can be included through simple, linear cost functions of the form 
\begin{equation}
    C(\gamma_1,\gamma_2;r) = \gamma_1 + r\gamma_2.
\end{equation}
Here, the parameter $r$ represents the \textit{relative cost} of preemptive or reactive suppression. At $r = 1$, the cost in each stage is equal. Taking $r < 1$, for example, could represent a decrease in suppression cost as mobilization of resources allows more cost-effective deployment. 

Here a risk-neutral utility function $U(\gamma_1,\gamma_2)$ is constructed by combining the probability of a bad outcome (no absorption, or escape) with the cost of suppression and a weighting parameter $q$:
\begin{equation}
    \label{eq:alloc_utility}
    -U(\gamma_1,\gamma_2) = \pr{\text{outcome}} + qC(\gamma_1,\gamma_2;r).
\end{equation}
The variable $q$ can be interpreted as an overall cost weight; as $q\to 0$, the utility is dominated by the probability of the bad outcome, while when $q \gg 1$ the utility is dominated by the cost of suppression. 

\begin{figure}[!h]
    \centering
    \includegraphics[width=0.9\columnwidth]{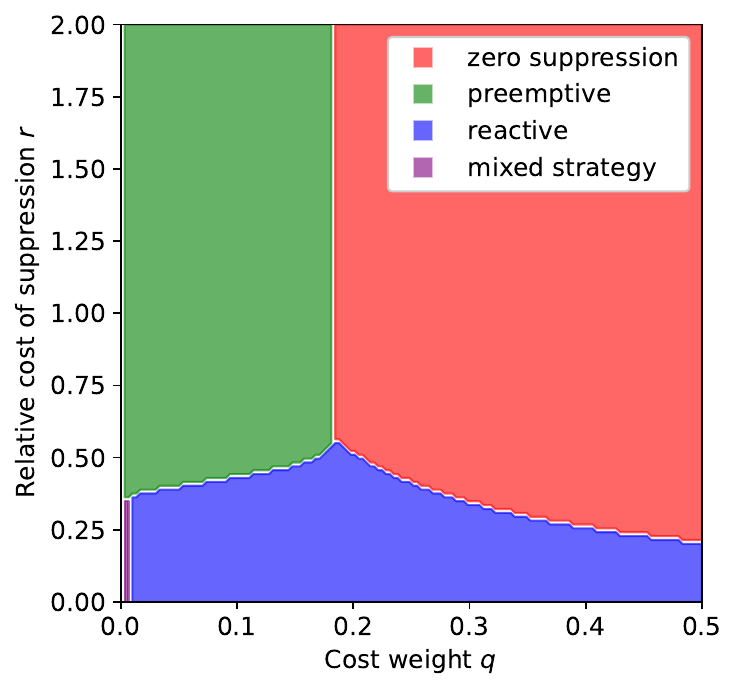}
    \caption{Minimizing Non-Absorption. Categories of optimal strategies which minimize the probability of no absorption plus a linear cost to suppression, parametrized by relative cost $r$ and cost weight $q$ as in \eqref{eq:alloc_utility}. In the majority of parameter space mixed strategies are not optimal, and at equal cost to suppression $r = 1$, preemptive strategies are completely favored.}
    \label{fig:alloc_abs_phases}
\end{figure}

\begin{figure}[!h]
    \centering
    \includegraphics[width=0.9\columnwidth]{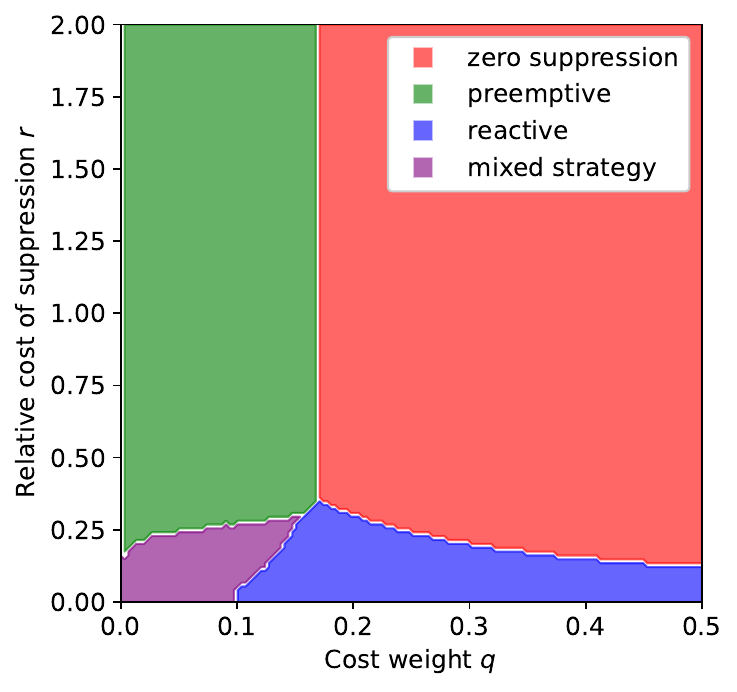}
    \caption{Minimizing Escape. Categories of optimal strategies which minimize the probability of escape $\pr{F(T_2) \geq 50}$ subject to cost as in Eq.~\eqref{eq:alloc_utility}. Compared to the phase diagram of Fig.~\ref{fig:alloc_abs_phases}, mixed strategies now form a larger share of the parameter space, but at $r = 1$ preemptive strategies are still preferred. }
    \label{fig:alloc_esc_phases}
\end{figure}

Optimal allocations $(\gamma_1^\star,\gamma_2^\star)$ are found by maximizing the utility $U(\gamma_1,\gamma_2)$. Depending on the values of the parameters $r,q$ in the utility function, optimal allocation strategies may prefer preemptive suppression $\gamma_1 > 0$, reactive suppression, $\gamma_2 > 0$, or a mix of the two. In Figs. \ref{fig:alloc_abs_phases}, \ref{fig:alloc_esc_phases} these differing strategy `phases' are color-coded as parameters $q,r$ in the cost function are varied. 


In a vast majority of cases, no mixed strategy is preferred. Indeed, at $r = 1$, where the cost of suppression in each stage is equal, a preemptive strategy is always preferred, subject to cost; see Fig.~\ref{fig:alloc_abs_phases}. Furthermore, the phases have extremely sharp boundaries: at low cost weight $q \approx 0.1$, strategies transition very quickly from preemptive to reactive as the relative cost parameter $r$ is modulated. These behaviors are also true when the outcome considered is escape, as shown in Fig.~\ref{fig:alloc_esc_phases}. 

When the bad outcome is escape (here with threshold $J = 50$), a larger regime is occupied by mixed strategies at low cost weight and low relative cost. However, in the regime of equal suppression cost in each stage $r = 1$, the conclusions are robust: preemptive strategies are preferred. Similar behavior is observed for non-linear cost functions such as $C(\gamma_1,\gamma_2) \propto (\gamma_1 + r\gamma_2)^\alpha$ for $\alpha > 1$. 

What does the existence of sharp phase boundaries say about general suppression strategies? In any situation, a mixed strategy results in a lower amount of suppression concentrated in a given stage. That these strategies are only preferred under low-cost conditions indicates that suppression is most effective when concentrated. In the present context, the most effective concentration is in the preemptive regime. 

The operational cost of fire suppression is a complex function of the existing suppression assets and their availability. It is unlikely that realistic cost functions are continuous: when suppression requirements necessitate a new modality of equipment, like aerial support, the cost and the available resources may increase sharply, with no ability to interpolate between low suppression (e.g., hand crew) regimes and high suppression (e.g., aerial tanker) regimes. Even if these details were incorporated into the utility function of Eq.~\eqref{eq:alloc_utility}, it would not change the effectiveness of preemptive versus reactive suppression.

\subsection{Allocations under uncertainty}
The effectiveness of preemptive strategies in mitigating bad outcomes subject to cost is not surprising: the longer a process evolves unencumbered, the more likely a catastrophic outcome becomes. Only when the cost of later suppression is very low do optimal strategies hold back initial suppression. A large attenuation in suppression cost over the course of a process is not a likely scenario in practice. However, in the situation just analyzed, allocations were made with perfect information about the future. There was no doubt if or when the wind event would occur, and therefore, no doubt that preemptive suppression would be both risk-reducing and cost-effective. 

In practice, when fires begin, not all available resources are immediately allocated to an initial attack. Some resources are withheld in order to respond to another potential fire. With the understanding that preemptive attack is crucial for keeping risk low, in the presence of uncertainty about future ignitions, optimal resource allocation to a given fire must hold back some suppression in order to respond quickly to another event, should it occur. 

\begin{figure}[!h]
    \centering
    \includegraphics[width=\columnwidth]{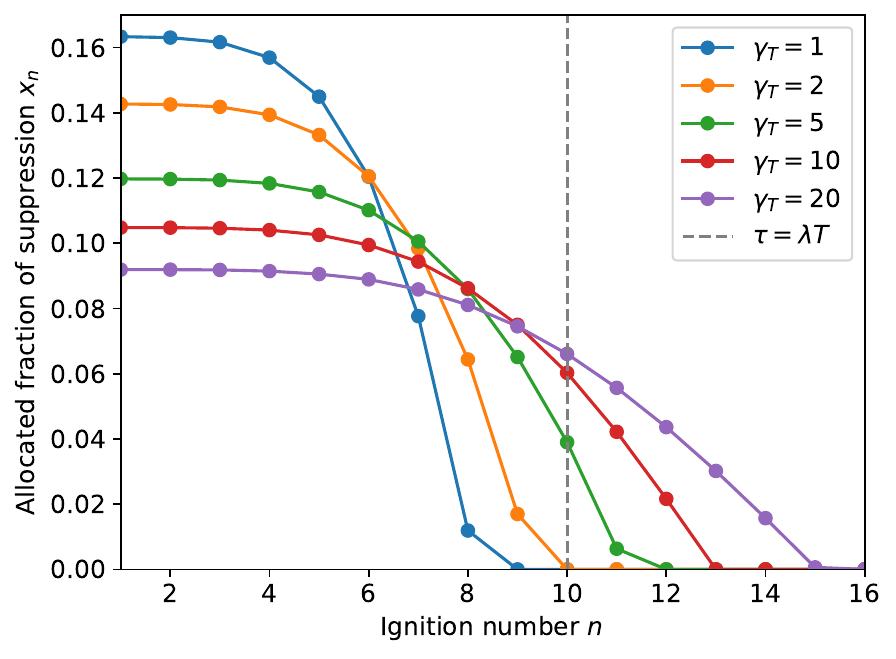}
    \caption{Optimal fractional suppression allocations for a series of potential ignitions. Fractional suppression allocations $x_n$ for the $n$-th ignition minimizing the expected number of escaped processes as in Eq.~\eqref{eq:uncertain_exp_escs}, with $J = 10,\ \tau = 10$. As the  amount of available suppression $\gamma_T$ increases, later events are allotted a greater fraction of suppression. }
    \label{fig:multi_event_alloc}
\end{figure}

A simple model using the birth-death-suppression process can be created to illustrate this logic. Imagine that some number $\tau = \lambda T$ of ignitions is expected to occur in the future. Specifically, let ignitions occur as a Poisson process with rate $\lambda$, and therefore $\lambda T$ ignitions are expected over a finite time period $T$. In this time period $T$, there is an amount $\gamma_T$ of suppression available to allocate to all ignitions. The probability of $n$ ignitions in this timeframe $T$ is 
\begin{equation}
    \pr{n \text{ ignitions}} = \frac{\tau^n}{n!}e^{-\tau}.
\end{equation}
Let each ignition nucleate a birth-death-suppression process with $N = 1,\ \beta = 1$. If a fraction $x_i$ of the available suppression $\gamma_T$ is allocated to the $i$-th event, the probability of escape with threshold $J$ is given by
\begin{equation}
    p_{esc}(x_i;\gamma_T,J) = \frac{\Gamma(J-1/2)\Gamma(1+x_i\gamma_T/2)}{\sqrt{\pi}\Gamma(J + x_i\gamma_T/2)}.
\end{equation}
In this setup, an allocation of resources is a choice of a function $x_n$ such that $\sum x_n = 1$: one chooses what fraction of available suppression $\gamma_T$ to allocate to the $n$-th event. The optimal allocation $x^\star_n$ should minimize the expected number of escaped processes over the timeframe $T$. This quantity can be expressed straightforwardly as
\begin{equation} 
    \label{eq:uncertain_exp_escs}
    \EE[\text{escapes in time $T$}](\mathbf{x}) = \sum_{n=1}^\infty \frac{\tau^n}{n!}e^{-\tau}\sum_{i = 1}^n p_{esc}(x_i).
\end{equation}
\begin{figure}[!h]
    \centering
    \includegraphics[width=\columnwidth]{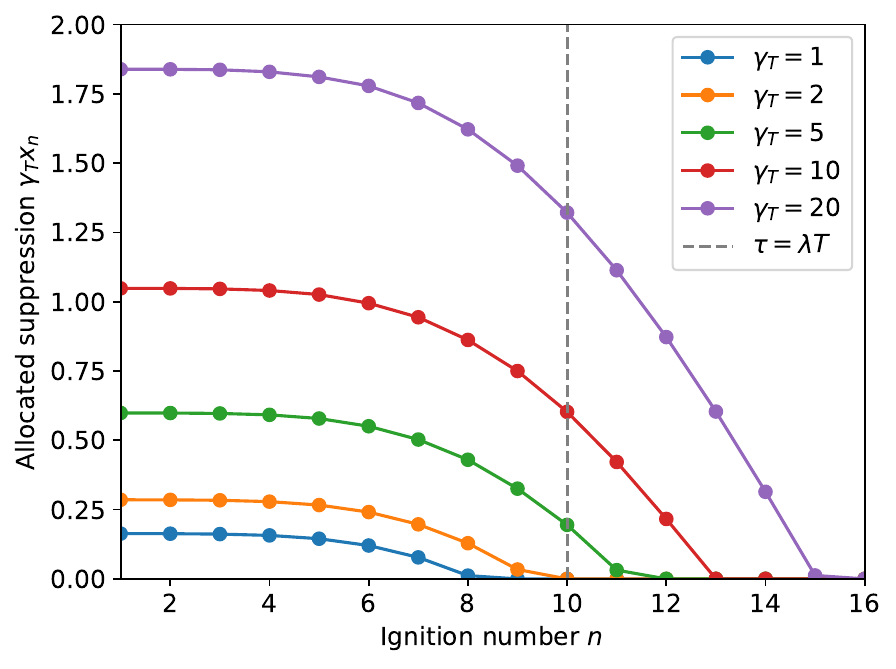}
    \caption{Optimal suppression allocations for a series of potential ignitions. Gross suppression allocations $\gamma_n = x_n\gamma_T$ for the $n$-th ignition are plotted with $\tau = 10,\ J = 10$ for different amounts of available suppression $\gamma_T$. While increased available suppression naturally allows more resources to be allocated to each event, the optimal allocation also extends these resources farther into the future. }
    \label{fig:multi_event_alloc_supp}
\end{figure}
By truncating this sum at a reasonably large value $n \gg \tau$, the constrained optimization problem of minimizing Eq. \ref{eq:uncertain_exp_escs}, subject to $\sum_n x_n = 1$, can be solved numerically by standard methods. The optimal gross allocations $\gamma_T x_n^\star$ are shown in Fig.~\ref{fig:multi_event_alloc_supp}, and the optimal allocation fraction $x^\star_n(\gamma_T)$ for various values of the available suppression $\gamma_T$ is shown in Fig.~\ref{fig:multi_event_alloc}.

The profiles of the optimal allocations $x^\star_n$ show that, as expected, uncertainty about future events causes some suppression to be held back from initial events despite high likelihood of their occurrence. That is, initial events receive most but not all available suppression. As the available suppression is increased, the optimal allocations save increasingly more resources for future events of relatively low likelihood. This remains true even as the suppression allocated to early events grows as the available suppression is increased. 

In realistic suppression scenarios, the true cost of suppression incorporates both the cost of deployment as well as the opportunity cost of allocating resources to one event under uncertainty about the occurrence of another, potentially more dangerous event. 

\section{trade-offs in suppression removal}
\label{sec:removal}
Having looked at allocations of two suppression rates over fixed time intervals, we now consider the `removal' problem, the third stage of the scenario illustrated in Fig.~\ref{fig:high_wind_schematic}. In the removal problem, a suppression rate $\gamma_3$ is chosen and applied until the chosen removal time $T_{rem}$, after which the process evolves freely. This represents the final stage ($t \geq T_2$) of the high wind scenario. Assuming that the process at time $T_2$ is in a state $(j_0,F_0)$ of definite population and footprint size, one must make a decision of a final allocation $\gamma_3$ and a duration $T \equiv T_{rem} - T_2$ over which to apply the suppression. The choice of allocation and duration will be informed by the proximity of $F_0$ to the escape threshold $J$, the probability of the population absorbing, and any cost considerations or constraints that may be present on the amount of available suppression $\gamma_3$. 

The fundamental tension addressed by this subproblem is that between fast, aggressive suppression strategies ($\gamma_3 \gg 1,\ T \ll 1$) and slow, sustained suppression strategies ($\gamma_3 \ll 1,\ T \gg 1$). By comparing the contours of constant asymptotic outcomes of the process, e.g. the average footprint $\ex{F(\infty)}$, one may find
a nonzero gradient in resource usage along the contours of constant outcome. That is, at a fixed value of the product $\gamma_3 T$, asymptotic outcomes may differ between the fast/aggressive suppression regime  and the slow/sustained suppression regime. 

Here, cost is assumed to depend directly on the total amount of suppression $\gamma_3 T$ and not the suppression rate or duration of suppression independently. The existence of a cost gradient, and hence a cost-optimal strategy, is completely independent of the specific cost function, and holds even if the set of allowed suppression values is restricted to a discrete subset. 

\subsection{Finding cost gradients}

\begin{figure}[!h]
    \centering
    \includegraphics[width=\columnwidth]{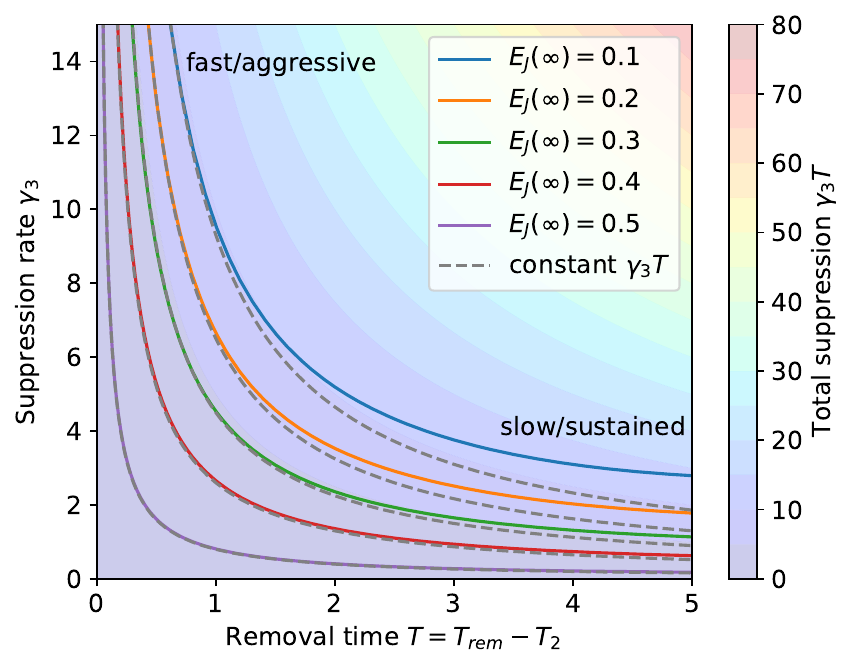}
    \caption{Trade-offs in escape probability. Contours of constant total suppression $\gamma_3 T$ (gray, dashed) overlaid on contours of constant asymptotic escape probability $E_J(\infty) \equiv \pr{F_\infty \geq J}$ in the $(\gamma_3,T)$ plane. At a fixed escape probability, strategies which apply a high suppression rate for a short amount of time are more cost-effective than applying a lower suppression rate over a longer period.}
    \label{fig:escape_removal_trade-off}
\end{figure}

Since in subcritical conditions $\beta < 1$ asymptotic absorption is a certainty, the focus in this section is on outcomes related to the footprint: the asymptotic probability of escape $E_J(\infty) = \pr{F_\infty \geq J}$ and the asymptotic average footprint $\ex{F(\infty)}$. The former is computed by the numerical strategy outlined in Eq.~\eqref{eq:asym_esc_recipe}, while the latter can be determined from the joint distribution $\PP_{j,F}(t)$ at the removal time $T_{rem}$. Recall that the average footprint $\ex{F(t)}$ solves
\begin{equation}
    \frac{d}{dt}\ex{F(t)} = \beta \ex{j(t)}.
\end{equation}
This equation can be integrated from $T_{rem}$ to asymptotic time, during which $\gamma = 0$, to obtain the following expression:
\begin{equation}
    \ex{F(\infty)} = \ex{F(T_{rem})} + \frac{\beta}{1 -\beta}\ex{j(T_{rem})}. 
\end{equation}
Both $E_J(\infty)$ and $\ex{F(\infty)}$ are computed across a range of suppression rates $\gamma_3$ and removal times $T_{rem}$, and the contour plots of these outcomes are compared to the contours of constant total suppression $\gamma_3 (T_{rem} - T_2) = \gamma_3 T$, as in Fig.~\ref{fig:escape_removal_trade-off}. 
We plot contours of constant total suppression (dashed) which coincide with contours of constant escape probability in the fast/aggressive regime (upper left). As these curves of constant total suppression travel into the slow/ sustained regime, it is clear that they fall below the contours of escape probability. The contours of constant escape probability move up the cost gradient (pictured as filled contours in the background of Fig.~\ref{fig:escape_removal_trade-off}) as they proceed into the slow-sustained regime.

\begin{figure}[!h]
    \centering
    \includegraphics[width=\columnwidth]{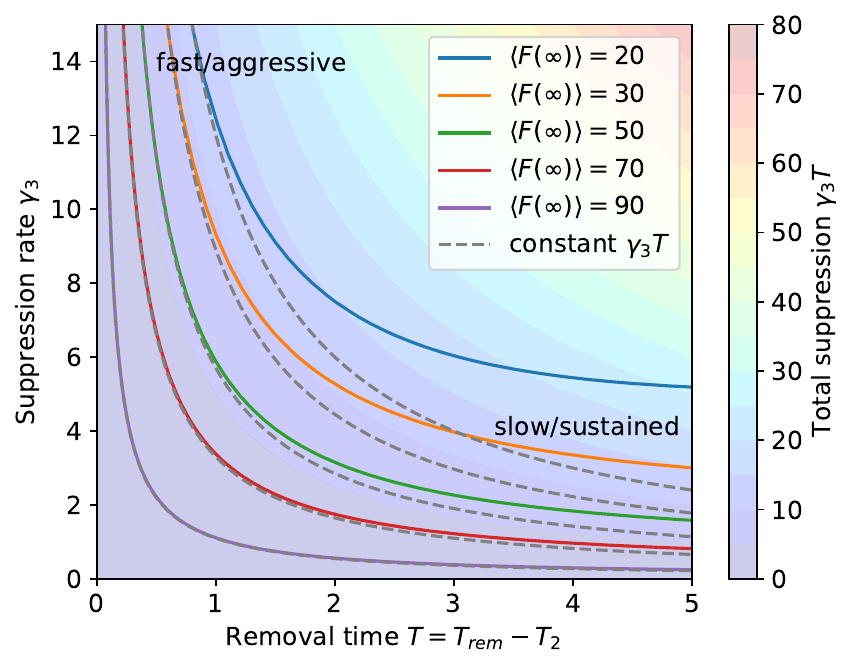}
    \caption{Trade-offs in final footprint. Contours of constant asymptotic average footprint with overlaid contours of constant total suppression $\gamma_3 T$ (gray, dashed). Fast/aggressive strategies (top left) are lower cost at fixed outcome compared to slow/sustained strategies (lower right); this cost gradient is more pronounced at smaller footprint values, corresponding to a lower likelihood of undesirable outcome.}
    \label{fig:footprint_trade-off}
\end{figure}

Said another way, there is a cost gradient along contours of constant escape probability from the slow/sustained regime (higher cost) to the fast/aggressive regime (lower cost). 
At high values of the escape probability, fast/aggressive and slow/sustained strategies appear to be approximately equivalent in terms of total suppression used. The cost gradient is strongest at low values of the asymptotic escape probability. These exact same conclusions are borne out when studying the contours of constant asymptotic average footprint, as in Fig.~\ref{fig:footprint_trade-off}.

This result is consistent with what was found in the allocation problem, where early, preemptive strategies are generally preferred and concentrated suppression is almost universally preferred. In the present case, the most resource-effective strategies concentrate suppression early and relax thereafter, rather than maintaining a consistent but lower level of suppression for an extended time. However, this effect is less pronounced for riskier outcomes: in Fig.~\ref{fig:footprint_trade-off}, the contours of constant average footprint at high values display much stronger cost-parity between the fast/aggressive and slow/sustained regimes. As one's risk tolerance decreases, concentrated strategies become more preferable. 

\section{Ignitions in dangerous conditions}
\label{sec:windignit}
The danger of high wind events is often that they both create ignitions and exacerbate fire spread once ignited. Previously, when considering ignitions in moderate conditions, allocations had been made with perfect knowledge of how future conditions would worsen. This is almost never the situation in practice. 

In this section we consider ignitions in dangerous conditions, corresponding to Case 2 in Fig.~\ref{fig:high_wind_schematic}. We directly address two crucial decisions relating to ignitions under high winds: pre-positioning and removal. By considering an ignition during the high wind event, we are effectively constructing a situation where an ignition is unexpected and there is no opportunity for preemptive suppression. We model the effect of pre-positioning by through resource delay: if suppression resources are effectively pre-positioned, there will be a decreased delay until they can effectively suppress the fire. The delay therefore serves as a temporal proxy for the spatial aspect of pre-positioning. 

\subsection{Pre-positioning and resource delay}
\begin{figure*}[t]
    \centering
    \includegraphics[width = \textwidth]{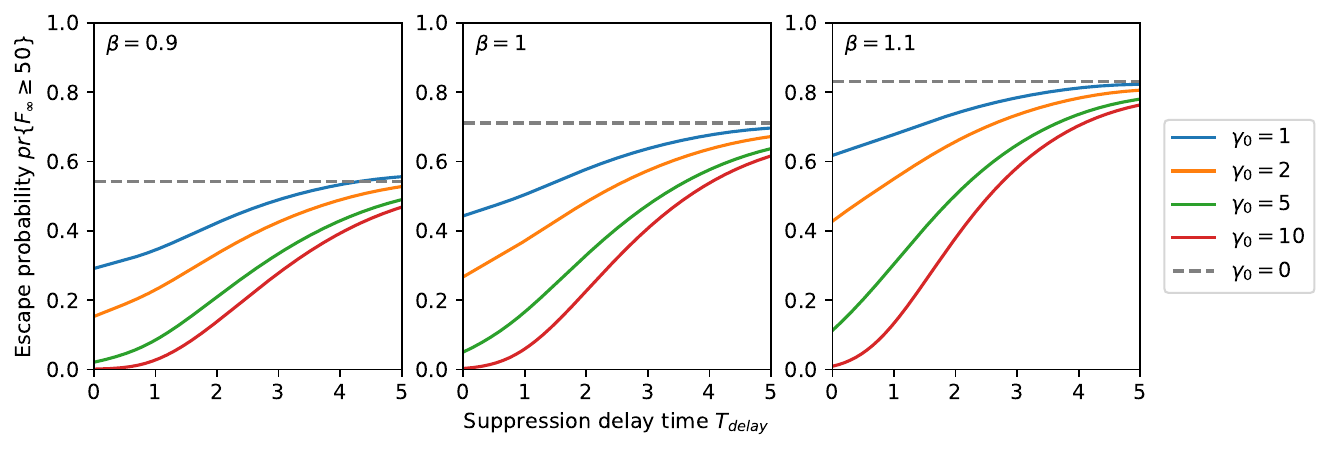}
    \caption{The importance of fast initial suppression. Effect of delayed suppression $\gamma_0$ on the escape probability $\pr{F_\infty \geq 50}$ for a process with $N = 10$, pictured for various values of $\gamma_0$ and $\beta = 0.9,\ 1,\ 1.1$. At long delays, even aggressive suppression only mildly lowers the probability of escape. Low, immediate suppression is more effective than high suppression brought in after a significant delay}
    \label{fig:delay}
\end{figure*}
Pre-positioning of emergency relief assets for disaster response is a well-known practice that is used globally by disaster management agencies. By pre-positioning resources, response time for potential events is lowered, but at a resource cost. In the context of wildfire, the phase of `initial attack,' or initial suppression, can be a deciding factor in whether a fire grows to extreme size and presents a danger to infrastructure and populations \cite{Rodrigues_Alcasena_Vega-Garcia_2019}. The problem of optimizing pre-positioning of resources based on cost and mobility constraints is inherently a spatial one and has been addressed by numerous works in the operations research literature \cite{suarez2017stochastic,seeberger2020new}. In the birth-death-suppression model, the focus is instead on the principal temporal aspect of pre-positioning: the effect of suppression delay on outcomes.

Specifically, we consider a birth-death-suppression process that begins with initial size $N = 10$ and has a constant birth rate. The process evolves freely until some time $T_{delay}$, after which a initial suppression rate $\gamma_0 > 0$ is introduced to the dynamics. The effect of the delayed arrival of initial suppression resources on the severity of the fire event is then quantified by the (asymptotic) escape probability $\pr{F_\infty \geq J}$ with an escape threshold of $J = 50$. This quantity is plotted in Fig.~\ref{fig:delay} versus the delay time $T_{delay}$ for different values of the birth rate representing mild, critical, and dangerous conditions. 

The results indicate that delay is more decisive than the rate of suppression: even if large amounts of suppression are deployed, if the delay is sufficiently long, there is a minor effect on the overall probability of escape as compared to the unsuppressed process.  In the intermediate delay regime, there is an approximately linear relationship between the delay and the probability of escape. Similar results were found in \cite{petrovic}, where after a large enough delay, the amount of suppression required to keep the escape probability constant increases exponentially.

In each plot of Fig.~\ref{fig:delay}, one can see that low, immediately deployed initial suppression, i.e. $\gamma_0 = 1$ at $T_{delay} = 0$, is more effective at reducing the probability of escape than substantial but highly delayed initial suppression, i.e. $\gamma_0 = 10$ at $T_{delay} = 5$. This observation has some operational ramifications. While heavy suppression always reduces risk, response time for initial attack can have an even greater effect than the magnitude of the suppression. This suggests that suppression resources should be geospatially distributed, thereby minimizing response time. This is consistent with both conventional wisdom and empirical studies that stress the importance of initial attack, particularly when ignitions occur in dangerous conditions. 

\subsection{Removal after a high wind ignition}
We now address a modified `removal' scenario. A process begins in dangerous conditions with $N = 10,\ \beta_0 = 1.2$, and proceeds with some low level of initial suppression $\gamma_0 = 2$. At time $T = 2$, the conditions relax to $\beta = 0.9$. A decision must then be made to apply a new suppression rate $\gamma_f$ until some removal time $T_{rem}$, which also functions as a decision parameter. There are a number of potential strategies. After the conditions relax, one could choose to decrease the suppression and allow the process to slowly extinguish; however, the rapid growth of the fire size during the high wind period increases the likelihood of escape, even in mild conditions. Alternatively, one could increase the suppression rate and aggressively suppress the fire, since the relaxed conditions mean higher suppression is now more effective. 

\begin{figure}[!h]
    \centering
    \includegraphics[width=\columnwidth]{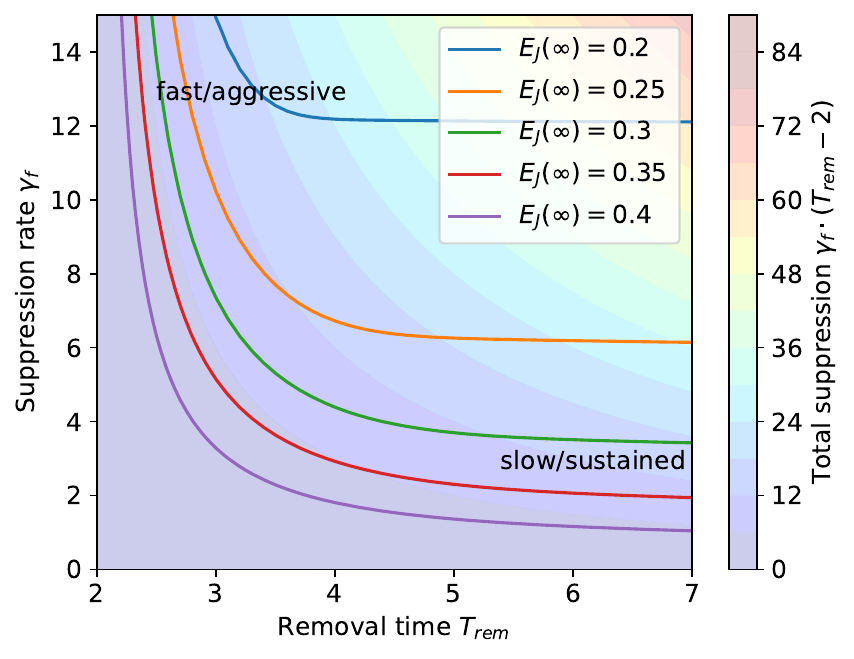}
    \caption{Trade-offs in escape probability after a high-wind ignition. Contours of constant escape probability $E_J(\infty) \equiv \pr{F_\infty \geq J}$ after an ignition in dangerous conditions superimposed over a contour map of the total suppression $\gamma_f (T_{rem} - 2)$; contours of constant total suppression are included in gray, dashed. At a fixed escape probability, the most cost-effective strategies clearly lie in the fast/aggressive regime in an even more pronounced fashion than Fig.~\ref{fig:escape_removal_trade-off}.}
    \label{fig:removal_highwind}
\end{figure}

The effect of differing removal strategies can be analyzed as in Sec. \ref{sec:removal}: by examining the contours of constant total suppression (after the initial attack) $\gamma_f\cdot (T_{rem}-2)$ and the contours of constant asymptotic escape $\pr{F_\infty \geq J}$. The contours of constant escape probability (with $J = 50$) are pictured in Fig.~\ref{fig:removal_highwind}, with a shaded map of the total suppression profile included in the background. 

The contours of fixed escape probability move quickly up the gradient of constant total suppression. In other words, after a high wind ignition, fast and aggressive strategies consume much less total suppression resources than slow and sustained strategies for a fixed outcome. The same conclusions were reached in Fig.~\ref{fig:escape_removal_trade-off}, but the cost gradients here are much more pronounced. 

While the effect of delay on outcomes suggests that resources should be distributed in order to decrease response time, the results in this subsection show the importance of maintaining the ability to deploy heavy, concentrated suppression. Temporally concentrated but very effective suppression is analogous to the use of aerial assets in wildfire response. Fast response even at low suppression rates is crucial at the beginning of a process. Once the process has grown, one needs the ability to deploy heavy, concentrated suppression. 

\section{Conclusion: optimal response in a high wind scenario}

The practice of wildfire suppression is performed by field personnel. People with years of experience, intuition for effective practices, and, in the modern era, a wealth of meteorological and geospatial data, inform the detailed motion of suppression resources around the spatial extent of an active wildfire. But, at a state or federal level, decisions relating to when and where to allocate a given amount of resources can greatly affect the cost of a fire event. Suppression expenditure has naturally grown with fire size and severity, a trend which only emphasizes the importance of effective resource usage and optimal allocation. 

In studying temporal resource allocations within the birth-death-suppression process, we see that distinct suppression strategies are regularly favored: concentrated, preemptive suppression is cost-optimal to lower the probability of bad outcomes. This is consistent with the widespread pre-positioning practices of many fire management agencies. In practice, the cost of pre-positioning must be balanced against the potential hazard of delayed resource mobilization, a hazard which we observed to depend strongly on the temporal mobilization delay. 

Tensions arise when operational constraints limit the ability to provide suppression or limit the available amount of suppression which can be deployed. From the `removal' problem, we saw that under total resource constraints, a limited amount of suppression should be temporally concentrated to have the greatest effect. 

The temporal trade-offs we observed have implications for the spatial distribution of resources and hence the optimization of response ability. Fast response and initial attack has a very strong effect on outcomes. Even a small amount of immediate suppression can be more valuable than delayed but heavy suppression. At the same time, once a fire event has begun, the optimal strategy is to deploy a temporally concentrated, high rate of suppression. This is similar to the usual practice of wildland firefighting, where initial attack is made by hand crews or small engines and, after some delay, aerial suppression assets provide intensely effective but temporally concentrated bouts of suppression: these strategies are reinforced by the birth-death-suppression model. 

While the focus of this paper is on the wildfire interpretation of the birth-death-suppression model, the process can also be used to describe the temporal evolution of any population of objects whose `death' rate is controlled by both natural effects and external suppression. As an example, the population $j(t)$ could represent the number of persons infected with a disease, and the suppression rate $\gamma$ would represent the effect of medical treatment or intervention. Much of the analysis carried out in this work could be adapted to address similar questions of the timing, scope, and duration of interventions for epidemic management. Birth-death Markov processes have historically been widely used for modeling the dynamics of disease spread \cite{kendall1949stochastic,Crawford2018}. 

Future work along the lines of this paper could analyze suppression allocation for multiple contemporaneous processes, extending the work of \cite{petrovic}. Integrating costs to resources and resource transfer, studying multi-event allocation would naturally address many of the tensions agencies are practically faced with in managing simultaneous fire events. Despite the lack of explicit spatial structure in the birth-death-suppression process, time delays and mobilization costs can implicitly model the effect of geospatially separated events. This model, and the general class of Markov birth-death processes, provide a simple but robust way to model the time evolution of spatial processes without being complicated by  detailed spatial dynamics.  

\acknowledgements

This work was supported in part by the Office of Secretary of Defense Strategic Environmental Research and Development Program (SERDP) Project No.~RC21-1233.

\appendix

\section{Cumulants of the birth-death-suppression process}
\label{sec:app_cumulants}
The connected cumulants of the population $j(t)$ and footprint $F(t)$ can be computed following the pioneering work of Kendall \cite{kendallGeneral}, who studied the cumulants of the zero-suppression linear birth-death process. The first step is to define the joint probability matrix $\PP(t)$ with elements
\begin{equation}
    \PP_{j,F}(t) = \pr{F(t) = F,\ j(t) = j}.
    \label{eq:jointProbMatDef}
\end{equation}
Recall that the footprint $F(t)$ counts the number of births, i.e. $F \to F + 1$ if and only if $j \to j + 1$. On general grounds, the joint probabilities $\PP_{j,F}(t)$ satisfy the dynamical (forward) equation
\begin{multline}
    \label{eq:jointProbdiffdiff}
    \frac{d}{dt}\mathcal{P}_{j, F}(t)=\lambda_{j-1} \mathcal{P}_{j-1, F-1}(t)+\mu_{j+1} \mathcal{P}_{j+1, F}(t)\\-\left(\lambda_j+\mu_j\right) \mathcal{P}_{j, F}(t).
\end{multline}
Presently, the aggregate birth/death rates $\lambda_j/\mu_j$ are given by 
\begin{equation}
    \lambda_j = \beta j; \quad \mu_j = j + \gamma; \quad \mu_0 = \lambda_{-1} = 0,
\end{equation}
with constant birth rate $\beta$ and suppression rate $\gamma$. In the case $\gamma = 0$, these reduce to the rates considered by Kendall.  However, for nonzero $\gamma > 0$, the behavior---and associated mathematics---changes considerably. This is principally because as one approaches the absorbing state $j \to 0$, the birth rate linearly decreases $\lambda \to 0$ while the death rate approaches a constant value $\mu \to \gamma$, and $\lim_{j\to 0}\mu_j \neq \mu_0$. This difference in limiting behavior causes the birth-death-suppression process to remain outside the universality class of absorbing processes to which the simple linear birth-death process belongs, as discussed in \cite{hathcock2022asymptotic,kessler2023extinction}. 

To determine the cumulants, begin by defining the joint generating function
\begin{equation}
    \Psi(t, z, w)=\sum_{j=0}^{\infty} \sum_{F=0}^{\infty} \PP_{j, F}(t) z^j w^F.
\end{equation}
The recurrence relation in Eq.~\eqref{eq:jointProbdiffdiff} implies that $\Psi(t,z,w)$ satisfies the inhomogeneous partial differential equation
\begin{multline}
    \frac{\partial \Psi}{\partial t}=\left(\beta z^2 w-z(\beta+1)+1\right) \frac{\partial \Psi}{\partial z}\\+\gamma\left(\frac{1-z}{z}\right)\left(\Psi(t, z, w)-f_0(t, w)\right),
    \label{eq:genPDE}
\end{multline}
where the function $f_0(t,w)$ is given by 
\begin{equation}
    f_0(t, w)=\sum_{F=0}^{\infty} P_{0, F}(t) w^F.
\end{equation}
As evident in Eq.~\eqref{eq:genPDE}, with $\gamma > 0$ the presence of the unknown function $f_0(t,w)$ renders a direct solution infeasible. Instead, consider the cumulant generating function $K(t,u,v) = \log \Psi(t,e^u,e^v)$. After the change of variables, one finds that $K(t,u,v)$ solves 
\begin{multline}
    \frac{\partial K}{\partial t}=\left(\beta e^{u+v}-(\beta+1)+e^{-u}\right) \frac{\partial K}{\partial u}\\+\gamma\left(e^{-u}-1\right)\left(1-e^{-K} f_0\left(t, e^v\right)\right). 
\end{multline}
To determine the lowest-order connected cumulants of $j(t),\ F(t)$, one solves the equation above order-by-order in $u,v$. By construction, the cumulant generating function $K(t,u,v)$ expands as
\begin{multline}
    K(t, u, v)=u\langle j(t)\rangle+v\langle F(t)\rangle+\frac{1}{2} u^2 \sigma_j^2(t)\\+u v \operatorname{Cov}(j, F)+\frac{1}{2} v^2 \sigma_F^2(t)+\cdots
\end{multline}
By equating terms at common orders in Eq.~\eqref{eq:genPDE}, one generates a hierachy of ordinary differential equations for each cumulant in the expansion above. Note that the function $f_0(t,e^v)$ expands as
\begin{equation}
    f_0(t,e^v) = \sum_{F=0}^\infty \PP_{0,F}(t) + v \sum_{F = 0}^\infty F\PP_{0,F}(t) + \OO(v^2).
\end{equation}
At order $\OO(u)$, one finds an equation for the average population $\ex{j(t)}$:
\begin{equation}
    \frac{d}{d t}\langle j(t)\rangle=(\beta-1)\langle j(t)\rangle-\gamma\left[1-p_A(t)\right].
\end{equation}
Here, the function $p_A(t) = \pr{j(t) = 0}$ is the absorption probability. It arises from the inhomogeneous term as
\begin{equation}
    f_0(t,1) = \sum_{F=0}^\infty \PP_{0,F}(t) = \pr{j(t) = 0} \equiv p_A(t). 
\end{equation}
In \cite{bds_wildfire}, we show that for processes with initial condition $j(0) = N$, the absorption probability is given by
\begin{equation}
    \label{eq:absorb_prob_ref}
    p_A(t) = \frac{B(z(t);N,\gamma+1)}{B(N,\gamma+1)};\quad z(t) = \frac{1-e^{(\beta-1) t}}{1-\beta e^{(\beta-1) t}},
\end{equation}
where $B(x;a,b)$ is the incomplete beta function. The absorption probability appears in the differential equation for the average population because absorbed processes contribute zero to the average over $j$, while the total probability mass across the states $j = 0,1,2,\ldots$ is conserved. Hence the decline in population due to the suppression factor must be discounted by what fraction of processes remain un-absorbed. 

Continuing to order $\OO(v)$, the average footprint $\ex{F(t)}$ satisfies
\begin{equation}
    \frac{d}{dt}\ex{F(t)} = \beta \ex{j(t)},
\end{equation}
expressing the intuitive fact that the footprint grows with the births of the population. One can exactly solve each of these equations and the corresponding equation for the population variance, but determining higher-order cumulants requires knowledge of the first order behavior of the function $f_0(t,e^v)$.

Below, we compute and present the integral expressions for the connected cumulants of the population $j(t)$ and footprint $F(t)$ in the birth-death-suppression process. We make no assumptions about the initial state of the process save that the cumulants may be computed in the initial state. 
\subsubsection*{Average population}
The average population solves the differential equation
\begin{equation}
    \frac{d}{d t}\langle j(t)\rangle=(\beta-1)\langle j(t)\rangle-\gamma\left[1-p_A(t)\right]. 
\end{equation}
The general solution is given by
\begin{multline}
    \langle j(t)\rangle=\ex{j(0)}e^{(\beta-1) t}\\-\gamma \int_0^t d \tau e^{(\beta-1) (t-\tau)}\left(1-p_A(t)\right),
\end{multline}
where the absorption probability is defined as $p_A(t) = \sum_j p_j(0)P_{j0}(t)$. In the case that the process begins in an initial state of definite population, one has $\ex{j(0)} = N$ for some integer $N$. In this case, we can write an exact expression for the integral in the above. Recall the following identity satisfied by the regularized incomplete beta function \cite{NIST:DLMF}:
\begin{equation}
    I_x(N, b)=1-\sum_{a=0}^{N-1} \frac{x^a(1-x)^b}{a B(a, b)} = \frac{B(x;a,b)}{B(a,b)},
\end{equation}
which holds for integral $N$ and arbitrary $b$. This identity, along with the expression for the absorption probability \eqref{eq:absorb_prob_ref} allows one to write
\begin{equation}
    \int_0^t e^{-(\beta-1) \tau}\left(1-p_A(\tau)\right) d \tau=\frac{1}{\gamma} \sum_{n=1}^N I_{z(t)}(n, \gamma),
\end{equation}
from which the exact expression
\begin{equation}
    \langle j(t)\rangle=e^{(\beta-1) t}\left[N-\sum_{n=1}^N I_{z(t)}(n, \gamma)\right]
\end{equation}
follows for integer $j(0) = N$. 
\subsubsection*{Average footprint}
The average footprint solves the equation
\begin{equation}
    \frac{d}{d t}\langle F(t)\rangle=\beta\langle j(t)\rangle,
\end{equation}
from which the solution may be directly integrated to yield
\begin{equation}
    \langle F(t)\rangle=\ex{F(0)}+\beta \int_0^t d \tau\langle j(\tau)\rangle.
\end{equation}
Since we can exactly forecast the average population across piecewise-constant birth and suppression rates, we can similarly exactly forecast the average footprint. 


\section{Glossary of formulae for the birth-death-suppression process}
\label{sec:app_glossary}
This section reproduces important formulae from our previous work \cite{bds_wildfire} which are referenced or used in the present analysis of the distribution of footprints. Throughout this section, the birth rate is $\beta$, the suppression rate is $\gamma$, and we work in time units where the death rate $\delta = 1$. To restore the formulae to physical units, one should make the substitutions $\beta \to \beta/\delta,\ \gamma \to \gamma/\delta,\ t \to t\delta$. 

To begin, define the auxiliary functions
\begin{equation}
    z(t)=\frac{1-e^{(\beta-1) t}}{1-\beta e^{(\beta-1) t}},
\end{equation}
and, with $s(t) = \exp{[(\beta-1)t]}$, 
\begin{equation}
    X(s)=\frac{(\beta s-1)(s-\beta)}{\beta(s-1)^2}.
\end{equation}
\subsubsection*{Distributions of the population}
The population $j(t)$ has transition matrix elements given by
\begin{multline}
P_{k \ell}(t)=  \frac{\pi_{\ell}}{\pi_k} \beta^k s^{\gamma+1} \\
 \times\left(\frac{1-\beta}{1-\beta s}\right)^{\gamma+2}\left(\frac{1-s}{1-\beta s}\right)^{k+\ell} F_{k \ell}(t)
\end{multline}
with 
\begin{multline}
F_{k \ell}(t)  =\frac{\Gamma(k+\ell+\gamma+2)}{\Gamma(\ell+1) \Gamma(k+\gamma+2)} \\
 \times{ }_2 F_1(-k,-\ell ;-1-k-\ell-\gamma ; X[s(t)]) .
\end{multline}
From this, the absorption probability is $p_A(t) = P_{N0}(t)$ if the population is in a definite state $j(0) = N$ at time $t = 0$. This function is succintly written as
\begin{equation}
    p_A(t)=\frac{B(z(t) ; N, \gamma+1)}{B(N, \gamma+1)},
\end{equation}
where $B(x;a,b)$ is the incomplete beta function. Its asymptotic limit may be computed as:
\begin{equation}
    p_A(\infty) = \begin{cases}
        1, & \beta \leq 1;\\
        \frac{B(1/\beta;N,\gamma +1)}{B(N,\gamma+1)}, & \beta > 1.
    \end{cases}
\end{equation}
\subsubsection*{Footprint orthogonal polynomials}
Computing the asymptotic footprint distribution and the corresponding asymptotic escape probability relies on a family of orthogonal polynomials $W_n(x)$ related by a change of variables to the Pollazcek polynomials \cite{van1990pollaczek}. We repeat these definitions here for reference. For more details on the Pollazcek polynomials and their properties, see \cite{koornwinder1989meixner,yermolayeva1999spectral,rui1996asymptotic,li2001asymptotics,araaya2004meixner}. 

We start by defining the following quantities:
\begin{gather}
    D(x)=\sqrt{x^2(\beta+1)^2-4 \beta},\\
    I_\beta=\frac{2 \sqrt{\beta}}{1+\beta}, \quad I_\beta \leq 1;\\
    u=\frac{x(\beta+1)}{2}+\frac{D(x)}{2}, \quad v=\frac{x(\beta+1)}{2}-\frac{D(x)}{2},\\
A=-\frac{\gamma+2}{2}-\frac{x \gamma(\beta-1)}{2 D(x)}, \\
B=-\frac{\gamma+2}{2}+\frac{x \gamma(\beta-1)}{2 D(x)}.
\end{gather}
The polynomials $W_n(x)$ may be evaluated as
\begin{multline}
W_n(x)= \frac{(\gamma+2)_n}{n!} u^{-n} \\
 \times{ }_2 F_1(-n,-B, \gamma+2 ;-u D(x) / \beta),
\end{multline}
although numerically, they are more efficiently evaluated using their recursion relation outlined in \cite{bds_wildfire}. 

The polynomials $W_n(x)$ are orthogonal with respect to a measure $d\sigma(x)$ with continuous and discrete support; the continuous portion $w(x) \equiv d\sigma/dx$ is given by 
\begin{multline}
    w(x)=\frac{\beta+\gamma+1}{2 \pi i} u^A v^B\\
    \times \frac{(u-v)^{-A}}{(v-u)^{B+1}} \frac{\Gamma(-A) \Gamma(-B)}{\Gamma(\gamma+2)}
\end{multline}
and supported on the interval $[-I_\beta,I_\beta]$. When $\beta > 1$ and $\gamma > 0$, the measure $d\sigma(x)$ is additionally supported on an infinite set of point $\pm x_k$ which satisfy
\begin{equation}
x_k^2=\frac{\beta(\gamma+2(k+1))^2}{(\gamma+(k+1)(\beta+1))(\beta \gamma+(k+1)(\beta+1))} .
\end{equation}
At the points $\pm x_k$ the measure has weight
\begin{multline}
    \Delta_k=\frac{\beta+\gamma+1}{\beta^{\gamma+3}}\left(u_k\right)^{-k}\left(v_k\right)^{k+\gamma+}\\
    \times \frac{(\gamma+2)_k}{k!} \cdot \frac{D_k^{\gamma+4}}{2 \gamma(\beta-1)},
\end{multline}
leading to the complete orthogonality relation
\begin{multline}
    \int_{-I_\beta}^{I_\beta}dx\ w(x) W_n(x) W_m(x) +\sum_{k=0}^{\infty} \Delta_k W_n(x_k) W_m(x_k)\\
    +\sum_{k=0}^{\infty} \Delta_k W_n(-x_k) W_m(-x_k)=h_n \delta_{n, m}.
\end{multline}
\subsubsection*{Asymptotic footprint distributions}
The footprint $F$ at absorption is linearly related to the initial population size $j(0) = N$ and the number of transitions $n_T$ that occurred in the lifetime of the absorbing process. The probability of a process absorbing with footprint $F$ is written as a spectral integral
\begin{multline}
    \label{eq:asymp_ftpt}
    \pr{F_\infty = F} = \frac{\gamma+1}{\beta+\gamma+1} \\\times \int_{-1}^1d \sigma(x)\ x^{2(F - F_0) + N - 1} W_{N-1}(x).
\end{multline}
Here $F_0 = F(0)$ is the initial footprint of the process and $d\sigma(x)$ is the complete measure of orthogonality, including both continuous and discrete elements as applicable to the phase of the process. 

To compute the probability of escape $E_J \equiv \pr{F_\infty \geq J}$ at some escape threshold $J$, one writes
\begin{equation}
    E_J = 1 - \sum_{F=F_0}^{J-1} \pr{F_\infty = F},
\end{equation}
subtracting from unity as only the complementary probability measure of $E_J$ is compactly supported. This quantity can be expressed directly in an integral form as
\begin{multline}
    \label{eq:escape_prob}
    E_J = 1 - \frac{\gamma+1}{\beta+\gamma+1}\int_{-1}^1 d\sigma(x)\\\times
    \frac{x^{N-1} - x^{2(J-F_0) + N - 1}}{1-x^2}W_{N-1}(x).
\end{multline}
These integrals may be performed numerically by standard quadrature methods. 

\bibliographystyle{unsrt}
\bibliography{refs}

\begin{thebibliography}{10}

\bibitem{Pandey_Huidobro_Lopes_Ganteaume_Ascoli_Colaco_Xanthopoulos_Giannaros_Gazzard_Boustras_etal_2023}
Pooja Pandey, Gabriela Huidobro, Luis~Filipe Lopes, Anne Ganteaume, Davide
  Ascoli, Conceição Colaco, Gavriil Xanthopoulos, Theodore~M. Giannaros, Rob
  Gazzard, Georgios Boustras, Toddi Steelman, Valerie Charlton, Euan Ferguson,
  Judith Kirschner, Kerryn Little, Cathelijne Stoof, William Nikolakis,
  Carmen~Rodriguez Fernández-Blanco, Claudio Ribotta, Hugo Lambrechts, Mariña
  Fernandez, and Simona Dossi.
\newblock A global outlook on increasing wildfire risk: Current policy
  situation and future pathways.
\newblock {\em Trees, Forests and People}, 14:100431, December 2023.

\bibitem{Burke_Driscoll_Heft-Neal_Xue_Burney_Wara_2021}
Marshall Burke, Anne Driscoll, Sam Heft-Neal, Jiani Xue, Jennifer Burney, and
  Michael Wara.
\newblock The changing risk and burden of wildfire in the {United} {States}.
\newblock {\em Proceedings of the National Academy of Sciences},
  118(2):e2011048118, January 2021.

\bibitem{Brown_Hanley_Mahesh_Reed_Strenfel_Davis_Kochanski_Clements_2023}
Patrick~T. Brown, Holt Hanley, Ankur Mahesh, Colorado Reed, Scott~J. Strenfel,
  Steven~J. Davis, Adam~K. Kochanski, and Craig~B. Clements.
\newblock Climate warming increases extreme daily wildfire growth risk in
  {California}.
\newblock {\em Nature}, 621(7980):760–766, September 2023.

\bibitem{MaryTaber_Elenz_Langowski_2013}
Mary~A. Taber, Lisa~M. Elenz, and Paul~G. Langowski.
\newblock {\em Decision Making for Wildfires: A Guide for Applying a Risk
  Management Process at the Incident Level}.
\newblock Number Gen. Tech. Rep. RMRS-GTR-298WWW. Fort Collins, CO, 2013.

\bibitem{jolly2015climate}
W~Matt Jolly, Mark~A Cochrane, Patrick~H Freeborn, Zachary~A Holden, Timothy~J
  Brown, Grant~J Williamson, and David~MJS Bowman.
\newblock Climate-induced variations in global wildfire danger from 1979 to
  2013.
\newblock {\em Nature communications}, 6(1):7537, 2015.

\bibitem{westerling2008climate}
Anthony~L Westerling and Benjamin~P Bryant.
\newblock Climate change and wildfire in {{California}}.
\newblock {\em Climatic Change}, 87:231--249, 2008.

\bibitem{liu2010trends}
Yongqiang Liu, John Stanturf, and Scott Goodrick.
\newblock Trends in global wildfire potential in a changing climate.
\newblock {\em Forest ecology and management}, 259(4):685--697, 2010.

\bibitem{abatzoglou2016impact}
John~T Abatzoglou and A~Park Williams.
\newblock Impact of anthropogenic climate change on wildfire across {Western}
  {US} forests.
\newblock {\em Proceedings of the National Academy of sciences},
  113(42):11770--11775, 2016.

\bibitem{gorte2013rising}
Ross Gorte and Headwater Economics.
\newblock {\em The rising cost of wildfire protection}.
\newblock Headwaters Economics Bozeman, MT, USA, 2013.

\bibitem{mattioli2022estimating}
W~Mattioli, C~Ferrara, E~Lombardo, Anna Barbati, L~Salvati, and A~Tomao.
\newblock Estimating wildfire suppression costs: a systematic review.
\newblock {\em International Forestry Review}, 24(1):15--29, 2022.

\bibitem{bayham2022economics}
Jude Bayham, Jonathan~K Yoder, Patricia~A Champ, and David~E Calkin.
\newblock The economics of wildfire in the {United} {States}.
\newblock {\em Annual Review of Resource Economics}, 14(1):379--401, 2022.

\bibitem{nifc_costs}
National Interagency~Fire Center.
\newblock Suppression costs.
\newblock
  \url{https://www.nifc.gov/fire-information/statistics/suppression-costs}.
  Last accessed 01 September 2024.

\bibitem{Swain_2021}
Daniel~L. Swain.
\newblock A shorter, sharper rainy season amplifies {{California}} wildfire
  risk.
\newblock {\em Geophysical Research Letters}, 48(5):e2021GL092843, March 2021.

\bibitem{Garner_Kovacik_2023}
Jonathan~M. Garner and Carly~E. Kovacik.
\newblock Extreme wildfire environments and their impacts occurring with
  offshore-directed winds across the {Pacific} {Coast} states.
\newblock {\em Weather, Climate, and Society}, 15(1):75–93, January 2023.

\bibitem{Guirguis_2023}
Kristen Guirguis, Alexander Gershunov, Benjamin Hatchett, Tamara Shulgina,
  Michael~J. DeFlorio, Aneesh~C. Subramanian, Janin Guzman-Morales, Rosana
  Aguilera, Rachel Clemesha, Thomas~W. Corringham, Luca Delle~Monache, David
  Reynolds, Alex Tardy, Ivory Small, and F.~Martin Ralph.
\newblock Winter wet–dry weather patterns driving atmospheric rivers and
  {Santa} {Ana} winds provide evidence for increasing wildfire hazard in
  {{California}}.
\newblock {\em Climate Dynamics}, 60(5–6):1729–1749, March 2023.

\bibitem{Abatzoglou_Kolden_Williams_Sadegh_Balch_Hall_2023}
J.~T. Abatzoglou, C.~A. Kolden, A.~P. Williams, M.~Sadegh, J.~K. Balch, and
  A.~Hall.
\newblock Downslope wind‐driven fires in the {Western} {United} {States}.
\newblock {\em Earth’s Future}, 11(5):e2022EF003471, May 2023.

\bibitem{Billmire_French_Loboda_Owen_Tyner_2014}
Michael Billmire, Nancy H.~F. French, Tatiana Loboda, R.~Chris Owen, and
  Marlene Tyner.
\newblock {Santa} {Ana} winds and predictors of wildfire progression in
  {Southern} {California}.
\newblock {\em International Journal of Wildland Fire}, 23(8):1119, 2014.

\bibitem{Moritz_Moody_Krawchuk_Hughes_Hall_2010}
Max~A. Moritz, Tadashi~J. Moody, Meg~A. Krawchuk, Mimi Hughes, and Alex Hall.
\newblock Spatial variation in extreme winds predicts large wildfire locations
  in chaparral ecosystems.
\newblock {\em Geophysical Research Letters}, 37(4):2009GL041735, February
  2010.

\bibitem{Westerling_Cayan_Brown_Hall_Riddle_2004}
Anthony~L. Westerling, Daniel~R. Cayan, Timothy~J. Brown, Beth~L. Hall, and
  Laurence~G. Riddle.
\newblock Climate, {Santa} {Ana} winds and autumn wildfires in {Southern}
  {California}.
\newblock {\em Eos, Transactions American Geophysical Union}, 85(31):289–296,
  August 2004.

\bibitem{McGinnis_Kessenich_Mearns_Cullen_Podschwit_Bukovsky_2023}
Seth McGinnis, Lee Kessenich, Linda Mearns, Alison Cullen, Harry Podschwit, and
  Melissa Bukovsky.
\newblock Future regional increases in simultaneous large {Western} {USA}
  wildfires.
\newblock {\em International Journal of Wildland Fire}, 32(9):1304–1314,
  August 2023.

\bibitem{prescribed_2023}
{U.S. Department of Agriculture, Forest Service}.
\newblock {\em National Prescribed Fire Resource Mobilization Strategy}.
\newblock Number FS-1216. June 2023.

\bibitem{NationalInteragencyFireCenter_2023}
National Interagency~Fire Center.
\newblock {\em National Interagency Predictive Services Handbook}.
\newblock 2023.

\bibitem{arnette2019risk}
Andrew~N Arnette and Christopher~W Zobel.
\newblock A risk-based approach to improving disaster relief asset
  pre-positioning.
\newblock {\em Production and Operations Management}, 28(2):457--478, 2019.

\bibitem{seeberger2020new}
Rachel~A Seeberger.
\newblock A new simulation-optimization model for wildland fire resource
  pre-positioning.
\newblock Master's thesis, Monterey, CA; Naval Postgraduate School, 2020.

\bibitem{suarez2017stochastic}
Daniel~Eduardo Su{\'a}rez~Bayona.
\newblock A stochastic programming approach for wildfire suppression:
  prepositioning and distribution of resources under uncertainty.
\newblock 2017.

\bibitem{petrovic}
Nada Petrovic, David~L Alderson, and Jean~M Carlson.
\newblock Dynamic resource allocation in disaster response: Tradeoffs in
  wildfire suppression.
\newblock {\em PloS one}, 7(4):e33285, 2012.

\bibitem{kendall1949stochastic}
David~G Kendall.
\newblock Stochastic processes and population growth.
\newblock {\em Journal of the Royal Statistical Society. Series B
  (Methodological)}, 11(2):230--282, 1949.

\bibitem{bds_wildfire}
George Hulsey, David~L. Alderson, and Jean Carlson.
\newblock Birth-death-suppression {M}arkov process and wildfires.
\newblock {\em Phys. Rev. E}, 109:014110, Jan 2024.

\bibitem{karlinear}
Samuel Karlin and James McGregor.
\newblock Linear growth, birth and death processes.
\newblock {\em Journal of Mathematics and Mechanics}, 7(4):643--662, 1958.

\bibitem{kendallGeneral}
David~G. Kendall.
\newblock On the generalized ``birth-and-death" process.
\newblock {\em The Annals of Mathematical Statistics}, 19(1):1--15, 1948.

\bibitem{karlinDiff}
S.~Karlin and J.~L. McGregor.
\newblock The differential equations of birth-and-death processes, and the
  {Stieltjes} moment problem.
\newblock {\em Transactions of the American Mathematical Society},
  85(2):489--546, 1957.

\bibitem{karlinClass}
Samuel Karlin and James McGregor.
\newblock The classification of birth and death processes.
\newblock {\em Transactions of the American Mathematical Society},
  86(2):366--400, 1957.

\bibitem{randomwalksKarlin}
Samuel Karlin and James McGregor.
\newblock {Random walks}.
\newblock {\em Illinois Journal of Mathematics}, 3(1):66 -- 81, 1959.

\bibitem{Moritz_Morais_Summerell_Carlson_Doyle_2005}
Max~A. Moritz, Marco~E. Morais, Lora~A. Summerell, J.~M. Carlson, and John
  Doyle.
\newblock Wildfires, complexity, and highly optimized tolerance.
\newblock {\em Proceedings of the National Academy of Sciences},
  102(50):17912–17917, December 2005.

\bibitem{Ryan_1991}
Gary Ryan.
\newblock {\em Sundowner Winds: A Report on Significant Warming Events
  Occurring in Santa Barbara, {California}.}
\newblock Santa Maria, CA, July 1991.

\bibitem{lam2015numba}
Siu~Kwan Lam, Antoine Pitrou, and Stanley Seibert.
\newblock Numba: A llvm-based python jit compiler.
\newblock In {\em Proceedings of the Second Workshop on the LLVM Compiler
  Infrastructure in HPC}, pages 1--6, 2015.

\bibitem{Rodrigues_Alcasena_Vega-Garcia_2019}
Marcos Rodrigues, Fermín Alcasena, and Cristina Vega-García.
\newblock Modeling initial attack success of wildfire suppression in
  {C}atalonia, {S}pain.
\newblock {\em Science of The Total Environment}, 666:915–927, May 2019.

\bibitem{Crawford2018}
Forrest~W. Crawford, Lam Si~Tung Ho, and Marc~A. Suchard.
\newblock Computational methods for birth-death processes.
\newblock {\em Wiley Interdisciplinary Reviews: Computational Statistics},
  10(2):e1423, March 2018.

\bibitem{hathcock2022asymptotic}
David Hathcock and Steven~H Strogatz.
\newblock Asymptotic absorption-time distributions in extinction-prone {Markov}
  processes.
\newblock {\em Physical Review Letters}, 128(21):218301, 2022.

\bibitem{kessler2023extinction}
David Kessler and Nadav~M. Shnerb.
\newblock Extinction time distributions of populations and genotypes.
\newblock 2023.

\bibitem{NIST:DLMF}
{\it NIST Digital Library of Mathematical Functions}.
\newblock \url{https://dlmf.nist.gov/}, Release 1.1.9 of 2023-03-15.
\newblock F.~W.~J. Olver, A.~B. {Olde Daalhuis}, D.~W. Lozier, B.~I. Schneider,
  R.~F. Boisvert, C.~W. Clark, B.~R. Miller, B.~V. Saunders, H.~S. Cohl, and
  M.~A. McClain, eds.

\bibitem{van1990pollaczek}
Walter Van~Assche.
\newblock Pollaczek polynomials and summability methods.
\newblock {\em Journal of mathematical analysis and applications},
  147(2):498--505, 1990.

\bibitem{koornwinder1989meixner}
Tom~H Koornwinder.
\newblock Meixner--{Pollaczek} polynomials and the {Heisenberg} algebra.
\newblock {\em Journal of mathematical physics}, 30(4):767--769, 1989.

\bibitem{yermolayeva1999spectral}
Oksana Yermolayeva and Alexei Zhedanov.
\newblock Spectral transformations and generalized {Pollaczek} polynomials.
\newblock {\em Methods and Applications of Analysis}, 6(3):261--280, 1999.

\bibitem{rui1996asymptotic}
Bo~Rui and R~Wong.
\newblock Asymptotic behavior of the {Pollaczek} polynomials and their zeros.
\newblock {\em Studies in Applied Mathematics}, 96(3):307--338, 1996.

\bibitem{li2001asymptotics}
X~Li and R~Wong.
\newblock On the asymptotics of the {Meixner}—{Pollaczek} polynomials and
  their zeros.
\newblock {\em Constructive approximation}, 17:59--90, 2001.

\bibitem{araaya2004meixner}
Tsehaye~K Araaya.
\newblock The {Meixner}--{Pollaczek} polynomials and a system of orthogonal
  polynomials in a strip.
\newblock {\em Journal of Computational and Applied Mathematics},
  170(2):241--254, 2004.

\end{thebibliography}

\end{document}